\documentclass[footinbib,a4paper,aps,prx,superscriptaddress,reprint,twocolumn,preprintnumbers,amsmath,amssymb,nobalancelastpage,10pt]{revtex4-2}

\usepackage{amsmath}
\usepackage{verbatim}
\usepackage{graphicx}
\usepackage{color}
\usepackage{tikz}
\usepackage{subfigure}
\usepackage{float}
\usepackage{slashed}
\usepackage{mathptmx}
\usepackage{amssymb}
\usepackage{amsmath}
\usepackage{amsfonts}
\usepackage{array}

\usepackage{bm}
\usepackage{xcolor}

\usepackage{braket}

\definecolor{mygrey}{gray}{0.35}
\definecolor{myblue}{rgb}{0.2,0.2,0.8}
\definecolor{mygreen}{rgb}{0.2,0.8,0.5}
\definecolor{myzard}{cmyk}{0,0,0.05,0}
\definecolor{mywhite}{rgb}{1,1,1}
\definecolor{myred}{rgb}{1,0.,0.3}

\usepackage[colorlinks=true,citecolor=myblue,linkcolor=myred]{hyperref}

 \def\ee{\mathord{\rm e}}
 
 \def\ii{\mathord{\rm i}}

\def\half{\textstyle\frac{1}{2}}

\renewcommand{\ii}{{\rm i}}
\renewcommand{\ee}{{\rm e}}
\def\beq{\begin{equation}}
\def\eeq{\end{equation}}
\def\barray{\begin{eqnarray}}
\def\earray{\end{eqnarray}}

\usepackage{latexsym,amsmath,amssymb,amstext}
\usepackage{graphicx}
\usepackage{dcolumn}
\usepackage{bm}
\usepackage[utf8]{inputenc}
\usepackage{amsmath, amsthm, amsfonts, amssymb}

\usepackage{physics}
\usepackage{dsfont}
\usepackage{graphicx}
\usepackage{cleveref} 
\usepackage{xfrac}
\usepackage{mathrsfs}
\usepackage[normalem]{ulem}
\usepackage{bbold}
\usepackage{soul}

\usepackage[noend]{algorithmic} 

\pdfoutput=1
\usepackage{bm} 
\usepackage{dcolumn} 
\usepackage{float}
\usepackage{graphicx} 
\usepackage{dsfont} 
\usepackage{xcolor}
\usepackage{physics}
\usepackage{comment}
\usepackage{soul}
\usepackage{enumitem}

\definecolor{mygrey}{gray}{0.35}
\definecolor{myblue}{rgb}{0.2,0.2,0.8}
\definecolor{mygreen}{rgb}{0.2,0.8,0.5}
\definecolor{myzard}{cmyk}{0,0,0.05,0}
\definecolor{mywhite}{rgb}{1,1,1}
\definecolor{myred}{rgb}{1,0.,0.3}

\usepackage[colorlinks=true,citecolor=myblue,linkcolor=myred]{hyperref}

\setlist[itemize]{leftmargin=1em}

\newcommand{\rwth}{Institute for Quantum Information, RWTH Aachen University, D-52056 Aachen, Germany}
\newcommand{\ucm}{Instituto de Física Teórica, UAM-CSIC, Universidad Autónoma de Madrid, Cantoblanco, 28049 Madrid, Spain}
\newcommand{\fzj}{Peter Grünberg Institute, Theoretical Nanoelectronics, Forschungszentrum Jülich, D-52425 Jülich, Germany}

\newcommand{\pd}[2]{\frac{\partial#1}{\partial#2}}

\definecolor{mygrey}{gray}{0.35}
\definecolor{myblue}{rgb}{0.2,0.2,0.8}
\definecolor{mygreen}{rgb}{0.2,0.8,0.5}
\definecolor{myzard}{cmyk}{0,0,0.05,0}
\definecolor{mywhite}{rgb}{1,1,1}
\definecolor{myred}{rgb}{1,0.,0.3}

\begin{document}

\title{Compressed-sensing Lindbladian quantum tomography with trapped ions}

\author{Dmitrii Dobrynin} \email{d.dobrynin@fz-juelich.de} \affiliation{\rwth} \affiliation{\fzj}
\author{Lorenzo Cardarelli} 
\affiliation{\rwth} \affiliation{\fzj}
\author{Markus Müller} 
\affiliation{\rwth} \affiliation{\fzj}
\author{Alejandro Bermudez} 
\affiliation{\ucm}


\begin{abstract}
 Characterizing the dynamics of quantum systems is a central task for the development of quantum information processors (QIPs). It serves to benchmark different  devices,  learn about their specific  noise, and  plan   
the next  hardware upgrades. However, this task is also very challenging, for it 
requires a large number of measurements and   time-consuming classical processing. 
 Moreover,  when interested in the  time dependence of the noise, there is an additional  overhead  
since the characterization  must be performed repeatedly within the time interval of interest. To overcome this limitation while, at the same time, ordering the learned  sources of noise by their  relevance, 
we  focus on the inference of the dynamical   generators of the noisy dynamics using Lindbladian quantum tomography (LQT). 
We propose two different improvements of LQT that alleviate  previous shortcomings. In the  weak-noise regime  of  current QIPs, 
we manage to linearize the maximum likelihood estimation of LQT, turning the constrained  optimization into a convex problem to reduce the classical computation cost  and  to improve its robustness. Moreover, by    introducing  compressed sensing techniques, we reduce the number of required measurements without sacrificing accuracy. 
To illustrate these improvements, we apply our LQT tools to trapped-ion experiments of single- and two-qubit gates, advancing in this way the  previous state of the art.
\end{abstract}

\maketitle

\setcounter{tocdepth}{0}
\begingroup
\hypersetup{linkcolor=black}
\tableofcontents
\endgroup

\section{\bf Introduction}
\label{sec:01_intro}

The progress on  quantum technologies witnessed over the past decades has relied on the development of high-precision techniques to  isolate,  manipulate, and interrogate quantum systems. When these systems increase in size, as required for instance in quantum-advantage demonstrations  based on quantum-information processors (QIPs)~\cite{10.5555/3135595.3135617,Harrow2017},  the 
development of efficient calibration and characterization strategies becomes a central task~\cite{Eisert2020,PRXQuantum.2.010102}.  This  requires a mix of tools to estimate  quantum states, their time evolution, and the operations used for their measurements, all of which are encompassed  within the broad subject of quantum tomography (QT). One may say that  QT is both a blessing and a curse, for it provides us with a well-defined route for the detailed learning of  quantum systems 
but, at the same time, it   entails a very large complexity. In most cases, this complexity scales with a certain power of the Hilbert space dimension $d=2^N$, and thus grows exponentially with   the number of qubits $N$. The specific so-called  sampling complexity of various QT  protocols is discussed in more detail in App.~\ref{app:QT_complexity}, which also serves to set our notation and introduce key  concepts in the QT literature. This includes rigorous  proofs that incorporate the limited  accuracy one can achieve in any realistic  system, including noise and the limited number of experimental runs to sample the probability distributions associated to the quantum state. 

In light of this exponential complexity,   QT in full generality has been mostly limited to small-sized   systems, as  illustrated in the
  estimation of quantum  states $\rho$~\cite{PhysRevA.40.2847,PhysRevA.55.R1561,Paris2004}. Small-size state-QT  has become a standard practice   in  various  platforms such as photons~\cite{PhysRevLett.70.1244,PhysRevA.64.052312}, neutral atoms~\cite{Janicke1995, Kurtsiefer1997,PhysRevLett.86.4721}, trapped ions~\cite{Poyatos1996, Leibfried1996,PhysRevLett.92.220402}, nuclear magnetic resonance~\cite{doi:10.1098/rspa.1998.0170,PhysRevLett.80.3408} and superconducting circuits~\cite{doi:10.1126/science.1130886,PhysRevLett.102.200402}. In an effort to minimise the complexity, alternative QT schemes have been proposed,  exploiting symmetry~\cite{dariano_23,PhysRevLett.105.250403,Moroder_2012} and entanglement~\cite{Cramer2010,PhysRevLett.111.020401,Lanyon2017}  arguments to constrain the possible states in order to reduce  the QT costs. A reasonable constraint in this respect 
  is that  states produced in high-fidelity QIPs  are close to  ideal pure states, and thus correspond to  low-rank density matrices $\rho$. This can be exploited via compressed sensing (CS) techniques, originally developed  for the recovery of sparse signals by random sampling with a   rate that is smaller than the one  expected from the signal bandwidth~\cite{Donoho2006, Candes2006}.   Even if  the resources of CS-QT still scale exponentially with the number of qubits $N$~\cite{PhysRevLett.105.150401,Flammia_2012,KUENG201788,7956181,7956181,comment} (see App.~\ref{app:QT_complexity}), the overall reduction can be important  in practice, and has allowed to push state-QT  to larger registers~\cite{,PhysRevLett.113.040503} including the experimental CS-QT of 7-qubit states~\cite{Riofrio2017}.

 QT requirements become even more demanding when one is not only interested in states, but also in their   dynamics $\rho(t_0)\mapsto\rho(t)=\mathcal{E}_{t,t_0}(\rho_0)$~\cite{doi:10.1080/09500349708231894,PhysRevLett.78.390,PhysRevA.63.020101,PhysRevA.63.054104,PhysRevA.68.012305}. As discussed in App.~\ref{app:QT_complexity}, even for a single snapshot of this dynamics at $t\in T=[t_0,t_{\rm f}]$, the complexity of this so-called process-QT presents an even faster exponential scaling with $N$. Therefore,  most experiments of full process-QT have also been  limited to small-sized  systems, such as  two-qubit entangling gates in nuclear magnetic resonance~\cite{Childs2001,10.1063/1.1785151}, photons~\cite{PhysRevLett.91.120402,PhysRevLett.93.080502}, trapped ions~\cite{PhysRevLett.97.220407,doi:10.1126/science.1177077,PhysRevLett.102.040501}, and superconducting circuits~\cite{Bialczak2010,PhysRevLett.109.240505}. Paralleling our discussion of possible strategies to reduce the cost of state-QT, one can restrict either the snapshot $\mathcal{E}_{t,t_0}$ to specific quantum channels of Pauli type~\cite{10.1145/3408039,fawzi2023lower}, or apply compressed-sensing techniques assuming the channel has  a reduced  Kraus rank  $r_{\kappa}\ll d^2=4^{N}$~\cite{Flammia_2012,Kliesch2019guaranteedrecovery}. Both of these techniques still have a complexity that scales exponentially with $N$ (see App.~\ref{app:QT_complexity}), although they can lead to a practical overall improvement. Note that in the context of high-fidelity QIPs, 
 low-rank channels  lie very close to  a specific target unitary 
operation, i.e. a quantum gate,  the knowledge of which  can be exploited to define a basis and reduce  the  resource scaling of process CS-QT  to a polynomial one~\cite{PhysRevLett.106.100401,Rodionov2014}.

Having discussed this, we can now delve into   the central theme of the current work. If one is interested in learning the noisy real-time dynamics of the QIP to estimate which is the optimal time duration of a gate, i.e.~the evolution time  for which errors are   minimized, the above process-QT of $\mathcal{E}_{t,t_0}$ require  repeating the whole procedure over and over again for each evolution time one is interested in $t\in T$. Since  the dynamics   of  closed quantum systems must be    generated by an underlying Hamiltonian $\mathcal{E}_{t,t_0}(\rho)=U(t,t_0)\rho U^\dagger(t,t_0)$, e.g. $U(t,t_0)={\rm exp}(-\ii(t-t_0)H)$ if the Hamiltonian is constant,  one  may sidestep  this  repetition overhead 
by  focusing on the estimation of the  Hamiltonian $H$~\cite{PhysRevA.69.050306,PhysRevA.71.062312,PhysRevA.73.052317,deClercq2016}. 
Using the $N$-qubit Pauli basis $E_\alpha\in\mathcal{B}_{\rm P}$~\eqref{eq:pauli_basis}, the Hamiltonian can be expressed in terms of  $d^2-1$ real numbers 
\beq
\label{eq:spanned_H}
H=\sum_{\alpha=1}^{d^2-1} c_\alpha E_\alpha,
\eeq
which can be grouped in a Hamiltonian vector   $\boldsymbol{c}\in\mathbb{R}^{d^2-1}$,  where we have excluded a trivial overall shift of the energies. Since $d^2=4^N$,  one may naively expect to face similar  exponential scalings of the complexity. However, when restricting the type of possible  Hamiltonians using  microscopic information~\cite{PhysRevLett.102.187203,Burgarth_2009,PhysRevLett.113.080401}, or exploiting entanglement arguments~\cite{PhysRevA.91.042129}, one can again reduce the complexity to a polynomial scaling. 
If such detailed  prior knowledge is not available, one can still exploit rather general  constraints on the locality  of the interactions in physical systems, and develop QT schemes    that  employ polynomial resources  to estimate the  Hamiltonian~\cite{PhysRevLett.107.210404,PhysRevX.8.031029,Qi2019determininglocal,PhysRevLett.122.020504,PhysRevLett.124.160502} (see App.~\ref{app:QT_complexity} for the description of  rigorous scalings).

From the perspective of compressed sensing, Hamiltonian tomography makes a drastic assumption by considering dynamical quantum maps $\mathcal{E}_{t,t_0}$ with  rank $r_{\kappa}=1$. In the context of  QIPs,  this amounts to limiting the possible  noise in gates  to a systematic  mis-calibration or drift leading  to coherent errors.  This type of errors does certainly not exhaust all important sources of noise in experiments, and one thus needs to go beyond this limit. For more generic errors,   the dynamical quantum map   has a larger rank and generally lacks an  inverse, falling into the class of completely-positive trace-preserving (CPTP) linear super-operators~\cite{nielsen00,watrous_2018}. 
In particular, there is a type of CPTP maps called Markovian, which can be divided  at any intermediate time $t'\in[t_0,t]$ as $\mathcal{E}_{t,t_0}=\mathcal{E}_{t,t'}\circ\mathcal{E}_{t',t_0}$, such  that $\mathcal{E}_{t,t'}$  is also a physical CPTP map~\cite{PhysRevLett.105.050403,Rivas_2014}. For the time-homogeneous CPTP maps of interest in our work,  this  divisibility follows  from the existence of a Liouvillian   generator~\cite{wolf2008}, 
 generalising the case of the Hamiltonian to  a so-called Lindbladian $\mathcal{E}_{t,t_0}(\rho)=\mathcal{E}_{t-t_0}(\rho)={\rm exp}\{(t-t_0)\mathcal{L}\}(\rho) $~\cite{Lindblad1976,Gorini1976}, acting on physical states as  
\begin{equation}
    \label{eq:lindblad_generator}
    \mathcal{L}(\rho) = -\ii \bigg[\sum_{\alpha}c_\alpha E_\alpha,\rho\bigg] + \sum_{\alpha,\beta}\! \frac{G_{\alpha\beta}}{2}\!\! \left( E_\alpha^{\phantom{\dagger}} \rho E_\beta^{\dagger} -  \big\{ E_\beta^{\dagger} E_\alpha^{\phantom{\dagger}}, \rho \big\} \!\right)\!.
\end{equation}
Here, we have introduced the dissipative Lindblad matrix $G\in\mathsf{Pos}(\mathbb{C}^{d^2-1})$,  which must be positive  semidefinite  to guarantee that the dynamical quantum map is indeed a one-parameter family of CPTP  channels~\cite{Breuer2002}. 
The  goal of {\it  Lindbladian  quantum tomography} (LQT) is to estimate the elements of the Lindblad matrix, possibly in conjunction with those of the Hamiltonian, from measurement data. As noted in App.~\ref{app:QT_complexity}, LQT has in principle the same number of parameters $d^2(d^2-1)$ to be learnt as full process-QT, with the advantage that it need not be repeated over time. Moreover, the knowledge of the Lindbladian  can give more physical insight onto the error sources and how to optimise their suppression, as compared to QT of  the full process matrix.

One may naively expect that estimating $\mathcal{E}_{t-t_0}$ via process-QT allows one to find the Lindbladian by taking the matrix logarithm, such that LQT would reduce to standard process-QT. However, as noted in~\cite{wolf2008,PhysRevA.67.042322,
Howard_2006,onorati2021fitting}, this requires searching through the complex logarithm branches, of which there is an infinite number, and can lead to inconsistencies in the presence of errors. LQT thus requires an independent tomographic strategy, the origin of which may be traced back to Ref.~\cite{PhysRevA.58.1723}, which put forth   a linear inversion method 
similar  to standard process-QT~\cite{doi:10.1080/09500349708231894,PhysRevLett.78.390}. In the work~\cite{Childs2001}, a microscopically-motivated parametrization of the dynamical quantum map was used for a more accurate  extraction of  the Lindblad generators from the linear inversion. 
However, in the presence of errors, the approximate inversion process can lead to unphysical estimates leading to generators that do not yield a CPTP map. In~\cite{PhysRevA.67.042322,Howard_2006}, a three-step procedure for Lindblad learning was proposed, which starts with a possible unphysical estimate, and then applies a nonlinear least-squares fit with an  added penalty for unphysical generators, followed by a final filtering step. 
Paralleling the advances in Hamiltonian learning, recent works have also   exploited the locality of interactions  to improve LQT~\cite{Bairey_2020,PRXQuantum.3.030324,frança2022efficient}, although we note that there are no rigorous proofs of the sampling complexity   to  our knowledge.

In this work, we unveil two directions of  improvement for LQT by 
constraining the estimation 
in a way that the associated dynamical map corresponds to an admissible physical process. The constraints are imposed via a  maximum-likelihood (ML)   philosophy,  as first considered in  the context of state-QT~\cite{PhysRevA.55.R1561} 
and process-QT~\cite{PhysRevA.63.020101,PhysRevA.63.054104,PhysRevA.68.012305}. 
For LQT~\cite{PhysRevApplied.18.064056,PhysRevA.101.062305}, ML estimation goes along  similar lines but, instead of using a generic CPTP dynamical quantum map $\rho(t)=\mathcal{E}_{t,t_0}(\rho_{0})$, it parametrizes the time evolution using  the Lindbladian $\rho(t)=\ee^{(t-t_0)\mathcal{L}}(\rho_{0})$~\eqref{eq:lindblad_generator}, and  incorporates it   in   a likelihood estimator. The estimation proceeds by  a non-linear optimization  subject to constraints on the Hermitian nature of the Hamiltonian $H$, and the positive semidefiniteness of the Lindblad matrix $G$. We also note that an alternative LQT strategy  has been presented in~\cite{PhysRevApplied.17.054018}, which minimizes a least-square estimator. These LQT  methods have been demonstrated in experiments with 
superconducting circuits~\cite{PhysRevApplied.18.064056,PhysRevApplied.17.054018} and 
 trapped ions~\cite{PhysRevA.101.062305}.

In this work, we partake in the development of  maximum-likelihood Lindbladian quantum tomography (ML-LQT), presenting advances that are then applied  to trapped-ion QIPs.
First, we show that a linearization procedure in the regime of low-error gates attained by 
modern  QIPs  transforms the ML-LQT estimator into a convex one which, in turn, allows for a  more efficient optimization. To make further improvements, we combine this linearized  ML-LQT  with compressed sensing, exploiting the fact that the  noise is not completely unstructured, but is instead controlled by a reduced number of leading noise sources that will depend on the specific QIP. The corresponding Lindblad matrix will be controlled by a few leading generators, which directly translates into its sparseness in that particular basis. We take advantage  of this feature by developing an accurate compressed sensing (CS)  estimation of the Lindbladian with a reduced number of measurements, demonstrating that informational completeness in  CS-LQT is not necessary.
We also introduce statistical convergence criteria 
to avoid over-fitting the measurement data.
Ultimately, we test the proposed  LQT methods on experimental data from  trapped-ion QIPs. This broadens the previous experimental implementations~\cite{PhysRevA.101.062305} by considering  two-qubit gates, preparing the ground for  more efficient  LQT schemes for  multi-qubit dynamics that will be explored in the future.

This article is organised as follows. In Sec.~\ref{sec:02_lindbladian_MLE}, we describe the maximum-likelihood estimation for LQT. Sec.~\ref{sec:03_linearization} contains our results for the linearization of ML-LQT for high-fidelity QIPs, together with two different algorithms for the conjugate gradient descent. We present a comparison of the linearized ML-LQT to the standard full ML-LQT, and identify in which regimes either of the two linearized algorithms has a better performance. In Sec.~\ref{sec:04_LLCS}, we introduce a compressed-sensing technique to improve the linearized LQT for situations in which the noise is structured and sparse, and one can estimate its generators with non informationally-complete datasets. These improvements of LQT by linearization and compressed sensing are applied to trapped-ion single- and two-qubit gates in Sec.~\ref{sec:06_ions}, advancing LQT to trapped-ion experiments with real not-injected noise. We present our conclusions and results in Sec.~\ref{sec:conclusions}. We include a more detailed discussion of the sampling complexity of quantum tomography in App.~\ref{app:QT_complexity}. App.~\ref{app:app_linearization} contains a detailed derivation of the linearizsation in the context of LQT, whereas App.~\ref{app:DIA_PGD} describe the details of  two algorithms for linearized ML-LQT.

\section{\bf  \! Maximum-likelihood Lindbladian quantum tomography~(LQT) }
\label{sec:02_lindbladian_MLE}

\begin{figure*}[!t]
  \centering
  \includegraphics[width=1\linewidth]{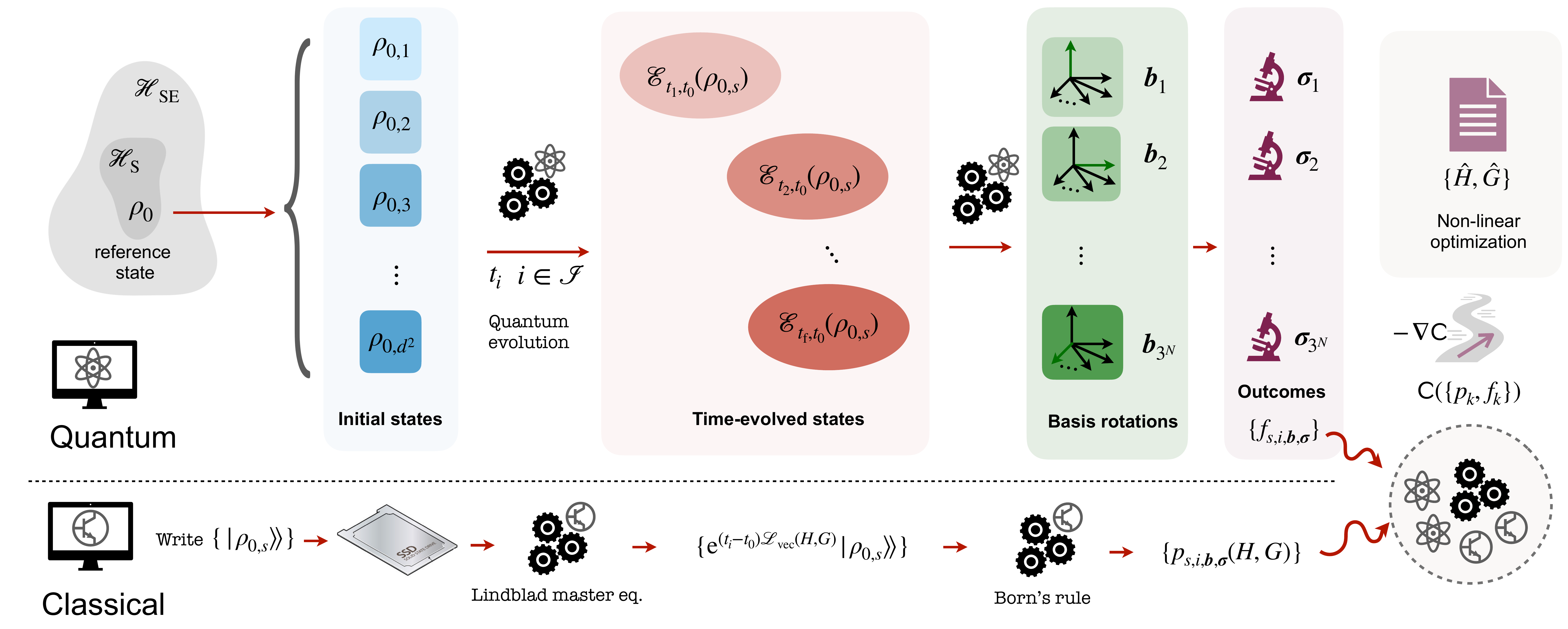}
  \caption{{\bf Scheme for a LQT protocol:} LQT is run both in the QIP and a classical computer. The former is initialized in a set of quantum states $\{\rho_{0,s},s\in\mathbb{S}_0\}\subset\mathsf{L}(\mathcal{H}_{\rm S})$ in a Hilbert space of dimension $d=dim(\mathcal{H}_{\rm S})$. These states evolve in time under the experimental CPTP map we want to estimate $\{\mathcal{E}_{t_i,t_0}(\rho_{0,s}): i\in\mathbb{I}_t\}$, which need not be unitary due to noise in the experimental controls, and the coupling to an ever-present environment $\mathcal{H}_{\rm S}\subset\mathcal{H}_{\rm SE}$. After the time evolution, the system is measured under a POVM $\{M_{\boldsymbol{b},\boldsymbol{m}_{\boldsymbol{b}}}: \boldsymbol{b}\in\mathbb{M}_{\boldsymbol{b}},\boldsymbol{m}_{\boldsymbol{b}}\in\mathbb{M}_{\boldsymbol{m}_{\boldsymbol{b}} }\}$. Each of these settings result in  specific outcomes  $\boldsymbol{m}_{\boldsymbol{b}}\in\mathbb{Z}_2^d$ that are arranged in relative frequencies $f_{s,i,\boldsymbol{b},\boldsymbol{m}_{\boldsymbol{b}}}$. The classical computer is used to find the corresponding probabilities $p_{s,i,\boldsymbol{b},\boldsymbol{m}_{\boldsymbol{b}}}(H,G)$ that are parametrized in terms of the system Hamiltonian $H$ and Lindblad matrix $G$ under the assumption that the CPTP map is well approximated by the Markovian semi-group associated to the  master equation generated by the Lindbladian~\eqref{eq:lindblad_generator}. Combining the finite frequency and the predicted probabilities, one can construct various possible estimators $\mathsf{C}(f_{s,i,\boldsymbol{b},\boldsymbol{m}_{\boldsymbol{b}}},p_{s,i,\boldsymbol{b},\boldsymbol{m}_{\boldsymbol{b}}})$, and solve a non-linear optimization problem using different gradient-descent strategies to obtain the estimates  $(\hat{H},\hat{G})$.}
  \label{fig:exp_setting}
\end{figure*}

In this section, we present a more detailed account of  ML-LQT,  starting from the theory of Markovian quantum master equations~\cite{10.1143/PTP.20.948,Zwanzig1960EnsembleMI,Breuer2002}. 
As noted in the introduction, for the dynamical quantum map $\mathcal{E}_{t,t_0}=\ee^{(t-t_0)\mathcal{L}}$ to represent a physically-admissible quantum evolution, the generator~\eqref{eq:lindblad_generator} must be expressed in terms of a positive semidefinite  Lindblad matrix $G$. Its diagonalization  $G\boldsymbol{u}_n=\gamma_n\boldsymbol{u}_n$ yields  non-negative eigenvalues $\gamma_n\geq 0$ 
  and associated  eigenvectors $\boldsymbol{u}_n=\sum_\alpha{{u}}_{\alpha, n}{\bf e}_\alpha$. This allows us to rewrite the  Liouvillian~\eqref{eq:lindblad_generator} in terms of the Lindblad-type master equation 
\begin{equation}
    \label{eq:lindblad_equation}
     \frac{{\rm d}\rho}{{\rm d}t} = -\ii [{H},\rho] + \sum_{n=1}^{d^2-1} \gamma_n \big( L_n \rho L_n^{\dagger} - {\half} \left\{ L_n^{\dagger} L_n, \rho \right\} \!\big),
\end{equation}
where $L_n=\sum_\alpha {u}_{\alpha, n}E_\alpha$ are the so-called jump operators, and $\gamma_n$ play the role of the corresponding decay rates~\cite{Breuer2002}.  In the case of atomic qubits, these jump operators are not arbitrary but typically correspond to leading noise sources, such as the amplitude-damping or dephasing processes~\cite{Chuang1997}. Estimating the decay rates and jump operators, especially when there are several noise sources present,  provides more direct insight than the full  dynamical quantum maps, as one can order the error generators $L_n$ in decreasing order of $\gamma_n$. Moreover, the knowledge of $\gamma_n,L_n$ allows one to assess which `software' and/or `hardware' improvements of the QIP would be most effective in combating the actual noise.

In order to learn the Lindbladian,  ML estimation introduces a cost function or estimator that quantifies the differences between the observed and  estimated probability distributions,
such as the negative log-likelihood estimator
\begin{equation}
    \label{eq:kullbackleibler}
	\mathsf{C}({p}_{1},{p}_{2}) = -\sum_k {p}_{1,k} \log({p}_{2,k}).
\end{equation}
 Here, the (multi)index $k$ spans the sample space of the probability distribution. In the context of process-QT, ${p}_{1}$ and ${p}_{2}$  correspond to the observed and predicted probability distributions for the measurements performed on the time-evolved state $\rho(t)=\mathcal{E}_{t,t_0}(\rho_0)$, respectively. 
The observed probabilities ${p}_{1,k}$ are approximated by  the relative measurement frequencies  ${f_k}$, the components of which are given by   the specific observed outcomes for each of the  LQT configurations composed of the initialization, evolution,  and measurement steps. 
Going beyond state QT and process QT, these relative frequencies are no longer a vector~\eqref{eq:state_QT_finite_fre} nor a matrix~\eqref{eq:process_QT_finite_fre}, but instead a tensor $f_{s,i,\mu}$ with indexes $k=(s, i, \mu )$. In this work, we consider  the positive operator-valued measure (POVM) elements $ \{M_\mu:\mu\in\mathbb{M}_f\}$ according to local Pauli measurements~\eqref{eq:local_pauli_proj}, which are applied to the time-evolved states from a set~\eqref{eq:IC_initial_set} of  initial states $\{\rho_{0,s}:s\in\mathbb{S}_0\}$ after a set of evolution times of interest $\{t_i:i\in\mathbb{I}_t\}\subset T$. We note that, in contrast to process-QT which typically focuses on a single channel at a single snapshot, the estimator~\eqref{eq:kullbackleibler} does in general include various evolution times~\cite{PhysRevApplied.18.064056}. As discussed in more detail below, 
we remark that these times need not densely cover  the interval of interest $|\mathbb{I}_t|\gg1$. In fact, for the Markovian evolutions hereby studied, it will suffice to use a single snapshot $|\mathbb{I}_t|=1$ to learn the Lindblad generators, whereas more snapshots will be required in situations in which the noise is time-correlated~\cite{santiago_in_prep}.  Once we obtain an estimate of the generators, it is possible to integrate the corresponding Lindblad master equation to infer the  dynamics at any desired time $t\in T$. 

The  estimator in Eq.~\eqref{eq:kullbackleibler} also depends on the predicted probabilities $p_2$, which are   derived from the solution of the  Lindblad equation~\eqref{eq:lindblad_equation} when considering the same set of POVM elements, initial states, and probing times. Hence, ${p}_{2,k}(\boldsymbol{c},G)={\rm Tr}\{M_\mu\ee^{(t_i-t_0)\mathcal{L}(\boldsymbol{c},G)}(\rho_{0,s})\}$, such that the log-likelihood estimator~\eqref{eq:kullbackleibler} is implicitly parametrized in terms of  the Hamiltonian and Lindblad matrix~\eqref{eq:lindblad_generator}. Due to the time-homogeneous character of the Markovian dynamical map, we can set $t_0=0$ without loss of generality, since there are no memory effects in the quantum evolution that depend on the specific initial time. From this perspective, the estimator~\eqref{eq:kullbackleibler} is related to the  likelihood function $\mathsf{L}(\{f_k,p_{2,k}\})$   for the joint multinomial probability distribution with which the   outcomes  would be observed with relative frequencies $\{f_k\}$, assuming an underlying statistical model that is parametrized by the Hamiltonian and Lindblad matrix $\{{p}_{2,k}(\boldsymbol{c},G)\}$. 
The maximum of this likelihood function gives the   model for which the observed outcomes are most probable, and corresponds a minimum after taking the negative logarithm and rescaling the result $\mathsf{C}(\{f_k,{p}_{2,k}(\boldsymbol{c},G)\})\propto-\log(\mathsf{L}(\{f_k,p_{2,k}\}))$ as in Eq.~\eqref{eq:kullbackleibler}.

The  ML-LQT protocol  is schematized  in Fig.~\ref{fig:exp_setting}, where we show  different triples $(\rho_{0,s}, t_i, M_\mu )$ that will be referred to as  configurations. Starting from a single reference state, a set of  initial states $\{\rho_{0,s},s\in\mathbb{S}_0\}$ is prepared by acting with local single-qubit gates, after which the system  evolves under the Lindbladian we aim at estimating for different times $\{t_i,i\in\mathbb{I}_t\}$, and is finally measured according to a POVM $\{M_\mu,\mu\in\mathbb{M}_f\}$. We assume for now on that one performs local Pauli measurements~\eqref{eq:local_pauli_proj}, such that $\mu=(\boldsymbol{b},\boldsymbol{m}_{\boldsymbol{b}})\in\mathbb{M}_f=\mathbb{M}_{\boldsymbol{b}}\times\mathbb{M}_{\boldsymbol{m}_{\boldsymbol{b}}}$, where the vector $\boldsymbol{b}$ specifies the $\{x,y,z\}$ basis for the measurement of each qubit, while the vector $\boldsymbol{m}_{\boldsymbol{b}}\in\{+,-\}^d$ determines the corresponding possible binary outcomes. Hence,  the  number of independent configurations~\eqref{eq:n_conf} is larger than the number of parameters  to be estimated per time step $n_{{\rm conf},i}=3^Nd^2(d-1)\geq d^2(d^2-1)$, such that the LQT scheme is informationally complete.
The  log-likelihood estimator  reads
\begin{equation}
	\label{eq:log_likelihood_Lindblad}
	\mathsf{C}_{\rm full}(\boldsymbol{c},G) = - \sum_{s,i,\mu} f_{s,i,\mu} \log({\rm Tr}\big\{M_\mu\ee^{(t_i-t_0)\mathcal{L}(\boldsymbol{c},G)}(\rho_{0,s})\big\}),
\end{equation}
where the relative-frequency tensor $f_{s,i,\mu}=N_{s,i,\boldsymbol{b},\boldsymbol{m}_{\boldsymbol{b}}}/N_{s,i,\boldsymbol{b}}$ 
is an approximation to the observed probability, calculated by   the ratio  of the number of observed  $\boldsymbol{m}_{\boldsymbol{b}}$-outcomes  $N_{s,i,\boldsymbol{b},\boldsymbol{m}_{\boldsymbol{b}}}$  with respect to  the   number of  measurements $N_{s,i,\boldsymbol{b}}=\sum_{\boldsymbol{m}_{\boldsymbol{b}}}N_{s,i,\boldsymbol{b},\boldsymbol{m}_{\boldsymbol{b}}}$ for a particular initial state, evolution time and measurement basis. 
Once the estimator is defined, its minimization subject to a positive semidefinite constraint  provides the full ML estimate of the Lindbladian
\beq
\label{LT_MLE}
\begin{split}
(\hat{\boldsymbol{c}}_{\rm full},\hat{G}_{\rm full})=& \texttt{ argmin}\big\{\mathsf{C}_{\rm full}(\boldsymbol{c},G)\big\}\\
&\texttt{ subject to}\;\; \;  \boldsymbol{c}\in\mathbb{R}^{d^2-1},\; G\in\mathsf{Pos}(\mathbb{C}^{d^2-1}).
\end{split}
\eeq
We note that the   positive semidefinite constraint can be imposed by  using a Cholesky decomposition $G=L^{\phantom{\dagger}}_{G}L_{G}^\dag$ in terms of  a lower triangular matrix $L_{G}$ with real and positive diagonal entries, paralleling some of the approaches for state-QT~\cite{PhysRevA.75.042108,Howard_2006}. As a result,
we can perform unconstrained numerical optimization using general purpose optimization libraries, such as the   C++ ALGLIB~\cite{Bochkanov}  for  nonlinear conjugate-gradient (CG) descent. 
For the descent iterations, we need to  numerically approximate  the gradient, which requires solving the Lindblad master equation~\eqref{eq:lindblad_generator} for sets of neighbouring $\boldsymbol{c},G$ parameters, which we accomplish by vectorizing the density matrix 	$\rho\to|\rho \rangle\!\!\rangle\in\mathbb{C}^{d^2}$, and calculating matrix exponentials of the corresponding Lindbladian. We will refer to this method  as  {full} ML-LQT~\cite{PhysRevApplied.18.064056,PhysRevA.101.062305} to distinguish it from the optimised routines for  linearized and compressed-sensing ML-LQT  that have been devised  in this work.

\section{\bf  Linearized LQT in high-fidelity quantum information processors} 
\label{sec:03_linearization}

Let us start by describing the first improvement for Lindblad tomography proposed in this work: linearization. The full ML-LQT~\eqref{LT_MLE}  is a non-convex optimization problem that can present  multiple local minima and saddle points, such that convergence to a global minimum is not guaranteed. As a consequence,   one may end up in a local minimum yielding a biased estimate of the Hamiltonian, dissipation rates and/or jump operators.
To overcome these limitations, we propose to use  
a linearization that can be applied in high-fidelity QIPs in which the coherent part of the time evolution is  known $U(t,t_0)={\rm exp}(-\ii (t-t_0) H)$, such that  $H$ can be excluded from the learning process. We can then restrict the estimation  procedure to the effect of the Lindblad matrix $G$  during a certain time scale $\Delta t$ that
 is small  even if the effect of the coherent evolution is not.
This is the case of gates in current QIPs, which are mildly affected by weak external noise.

With these assumptions, we can apply a  linearization procedure based on error process matrices~\cite{Korotkov2013} to the current ML-LQT, and show that the constrained minimization problem~\eqref{LT_MLE} turns into a convex one.
In this regime, as described in detail in App.~\ref{app:app_linearization}, the parametrized probabilities can be written as linear functions of the Lindblad matrix
\beq
\label{eq:est_prob_lin}
p^{\phantom{U}}_{s,i,\mu} = p^{\rm u}_{s,i,\mu} + \sum_{\alpha,\beta}\Phi_{s,i,\mu}^{pq} G_{pq},
\eeq
where  the contribution from the ideal gate unitary reads 
\beq
\label{eq:p_U}
p^{\rm u}_{s,i,\mu} = \text{Tr}\left\{M_\mu U(t_i,t_0)\rho_{0,s} U^\dag\!(t_i,t_0) \right\}.
\eeq
The weak Markovian noise linear in $G$ that we aim at  learning is fully contained in the second term, which depends on the following set of matrices $\{\Phi_{s,i,\mu}\}$ for each configuration, which are expressed  similarly   to Eq.~\eqref{eq:p_U}, namely
\beq
\label{eq:lin_contrib_noise}
\Phi_{s,i,\mu}^{pq}=\text{Tr}\left\{M_\mu U(t_i,t_0)\delta\rho^{pq}_s(t_i,t_0) U^\dag\!(t_i,t_0)  \right\}.
\eeq
The specific effect of the noise  is thus  contained in 
\beq
\label{eq:delta_rho_lin}
\delta\rho^{pq}_s(t_i,t_0)=\sum_{\alpha\beta}\int_{t_0}^{t_i}\!\!{\rm d} t'\!\left[\mathbb{W}^\dag\!( t') \mathbb{B}^{pq} \mathbb{W}( t')\right]_{\alpha\beta}\!
	E_\alpha^{\phantom{\dagger}}\rho_{0,s}E_\beta^\dag,
\eeq
which depends on  the matrices  $\{\mathbb{B}^{pq}\}$ and $\mathbb{W}(t')$ defined in Eqs.~\eqref{B_tensor} and~\eqref{eq:W_app}.
We refer the reader to the Appendix \ref{app:app_linearization} for the the full derivation, which starts with a Lindblad representation in an operator basis different from Eq.~\eqref{eq:lindblad_generator} in order to prepare the ground for the next compressed-sensing improvement.
The linearized  log-likelihood  estimator obtained after this weak-noise approximation reads as follows 
\begin{equation}
  \mathsf{C}_{\rm lin}(G) = -\sum_{s,i, \mu}f_{s,i,\mu}\log{\bigg( p^{\rm u}_{s,i,\mu} + \sum_{p,q}\Phi_{s,i,\mu}^{pq} G_{pq} \bigg)}.
\label{eq:linearized_loglikelihood}
\end{equation}

Hereafter, the method based on Eq. \eqref{eq:linearized_loglikelihood} will be referred to as {linear} ML-LQT. Let us comment on the relevance of this linearization for the   optimization. In the full ML-LQT~\eqref{eq:log_likelihood_Lindblad}, when considering a complete knowledge of the Hamiltonian, one must numerically integrate  the full Lindblad master equation for each time $t_i$, given an initial guess of the Lindblad matrix $\hat{G}_{\rm full}$, and  update it according to a gradient-descent method in the search for the optimal solution to the non-convex minimization of Eq.~\eqref{LT_MLE}. In the  linearized estimator~\eqref{eq:linearized_loglikelihood}, the  evaluation of Eq.~\eqref{eq:lin_contrib_noise} still requires a numerical integration  at each $t_i$, but this is highly simplified as we only need to exponentiate the Hamiltonian for an infinitesimal time step once, and not repeat the exponentiation of the vectorised Lindbladian for each updated value of $\hat{G}_{\rm lin}$. Moreover, the important advantage of the linearized ML-LQT is that the optimization problem  
\beq
\label{eq:lin_ML}
\begin{split}
\hat{G}_{\rm lin}=&\texttt{ argmin}\{\mathsf{C}_{\rm lin}(G)\}\\
&\texttt{ subject to }G\in\mathsf{Pos}(\mathbb{C}^{d^2-1}),
\end{split}
\eeq
has become convex, as the estimator is now a linear  function of $G$, and the positive semidefinite constraint draws a convex cone over the  Lindblad matrices. Linearized ML-LQT  is thus guaranteed to have a unique solution  $\hat{G}_{\rm lin}$ in contrast to $\hat{G}_{\rm full}$, which can have a more complex minima landscape. 

In App.~\ref{app:DIA_PGD},  we present two efficient methods for the convex minimization of the linearized ML-LQT estimator~\eqref{eq:linearized_loglikelihood}, namely the  {diluted iterative algorithm} (DIA) and the projected gradient descent with momentum (pGDM). Both of these methods have been considered in the context of state-QT~\cite{Rehacek2007,Bolduc2017} but, to our knowledge,  not for   Lindbladian quantum tomography. We start by comparing the generic conjugate gradient (CG) methods for the full ML-LQT~\eqref{LT_MLE} to the DIA approach for the linearized ML-LQT~\eqref{eq:lin_ML}, which
makes use of the analytical expression for the gradient of the linear estimator $\mathsf{C}_{\rm lin}$~\eqref{eq:linearized_loglikelihood}, 
together with a line search and  a conjugate-gradient descent, to converge towards the  unique global minimum $\hat{G}_{\rm lin}^{\rm DIA}$.
We consider a two-qubit system $N=2,d=4$, subject to a small Markovian
noise  corresponding to a random Lindblad matrix $G_{\rm true}$ obtained  by sampling a uniform random unitary in the Haar measure sense~\cite{mezzadri2007generate}. This unitary is applied  to an arbitrary state $\ket{\psi}\in\mathbb{C}^{15}\otimes\mathbb{C}^{15}$, after which one traces over one of the subsystems.  As a result, we  get a random positive semidefinite matrix  $G_{\rm true}\in\mathsf{Pos}(\mathbb{C}^{15})$ of trace one, sampled uniformly according to the Hilbert-Schmidt distance~\cite{Zyczkowski_2001}. To change the scale of $G_{\rm true}$, we  multiply the result by any desired prefactor, which is here set to  $({\rm Tr}\{G_{\rm true}\}) t_{\rm f}=0.25$. 
The Hamiltonian $H_{\rm true}$ for each generated $G_{\rm{true}}$ is chosen to produce a $\pi/2$ single qubit rotation along a random axis for each run.
Once a true  Lindbladian is randomly chosen, we simulate numerically the dynamics without any approximations, and obtain the exact POVM probabilities $f_{s,i,\mu}\mapsto p_{s,i,\mu}$, focusing on a single snapshot $t_i=t_{\rm f}$. By using the exact probabilities,  we momentarily dispense with  the effects of shot noise  in order to compare the convergence of the full and linearized ML-LQT approaches. In sections below, we will go beyond these approximations and also consider real experimental data with shot noise and other SPAM errors.

\begin{figure}[t]
  \centering
\includegraphics[width=0.8\linewidth]{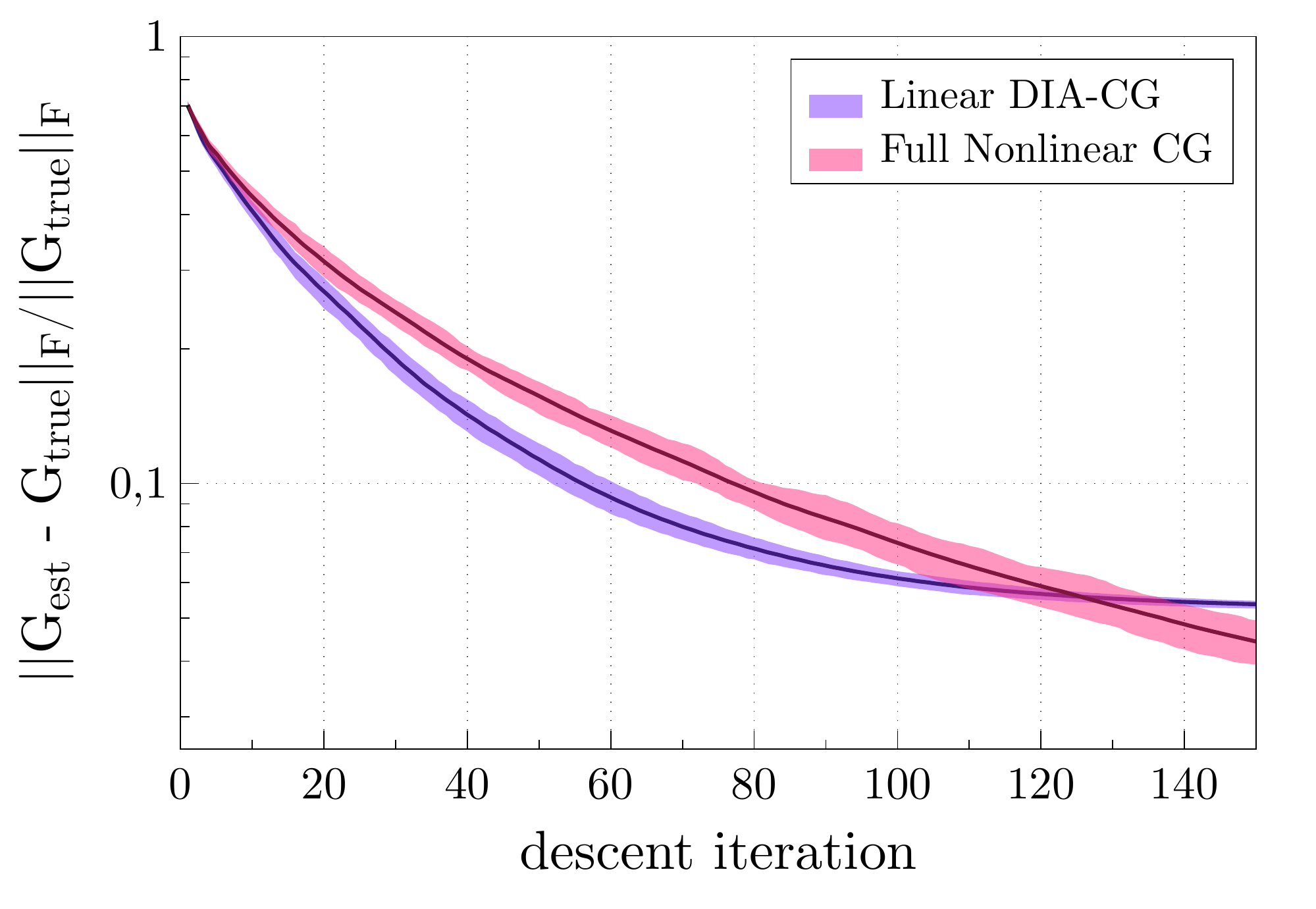}
    \caption{{\bf Performance comparison of full and linear  ML-LQT:}. We use
 the DIA algorithm for the linearized ML-LQT ~\eqref{eq:lin_ML}, discussed in App.~\ref{app:DIA_PGD}, and a general-purpose CG method from the {ALGLIB} library for the full ML-LQT~\eqref{LT_MLE}. We uniformly draw  $100$ semidefinite positive matrices that play the role of a true Lindblad matrix $G_{\rm true}\in\mathbb{C}^{15}\otimes\mathbb{C}^{15}$ with~$(\text{Tr}\{G_{\rm true}\})t_{\rm f} = 0.25$. We consider a single measuring time $|\mathbb{I}_{t}|=1$ in $T=[0,t_{\rm t}]$, which we take to be $t_i=t_f$, and $n_{\rm conf,i}=432$ configurations to achieve information completeness. The  true Lindblad matrix is    used to numerically generate the exact probability distribution $p_{s,i,\mu}(G_{\rm true})={\rm Tr}\big\{M_\mu\ee^{(t_i-t_0)\mathcal{L}(\boldsymbol{0},G_{\rm true})}(\rho_{0,s})$. We start by focusing on the asymptotic limit in which shot noise is absent, and directly use the exact $p_{s,i,\mu}(G_{\rm true})$ instead of the finite-shot relative frequencies $f_{s,i,\mu}$. We run the constrained minimization for the linear and full ML-LQT, finding the corresponding estimates $\hat{G}_{\rm lin}$ (purple) and $\hat{G}_{\rm full}$ (pink) for each of the random choices of the Lindblad matrix. We quantify the precision of the estimates by calculating the Frobenius norm of the difference $||\hat{G}-G_{\rm true}||_{\rm F}$, properly normalized, where we recall that this norm is the square root of the sum of the squares of all the matrix entries. In the figure we represent the mean and  20/80 percentiles for the distribution of this quantity among all the random draws of the Lindblad dynamics. 
   \label{fig:DIA_vs_ALG}}
\end{figure}

\begin{figure}[t]
    \centering
    \includegraphics[width=.8\linewidth]{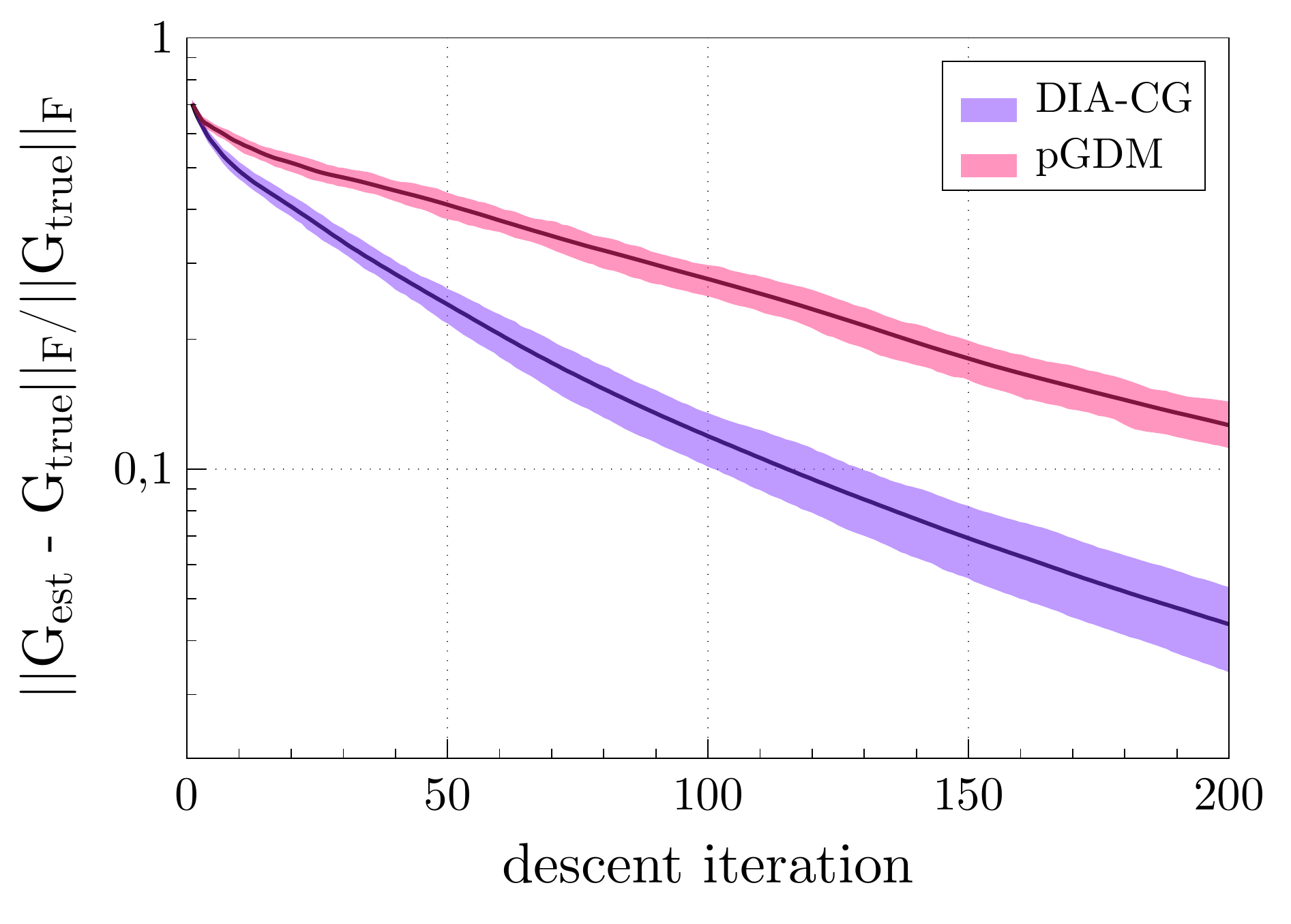}
    \includegraphics[width=.8\linewidth]{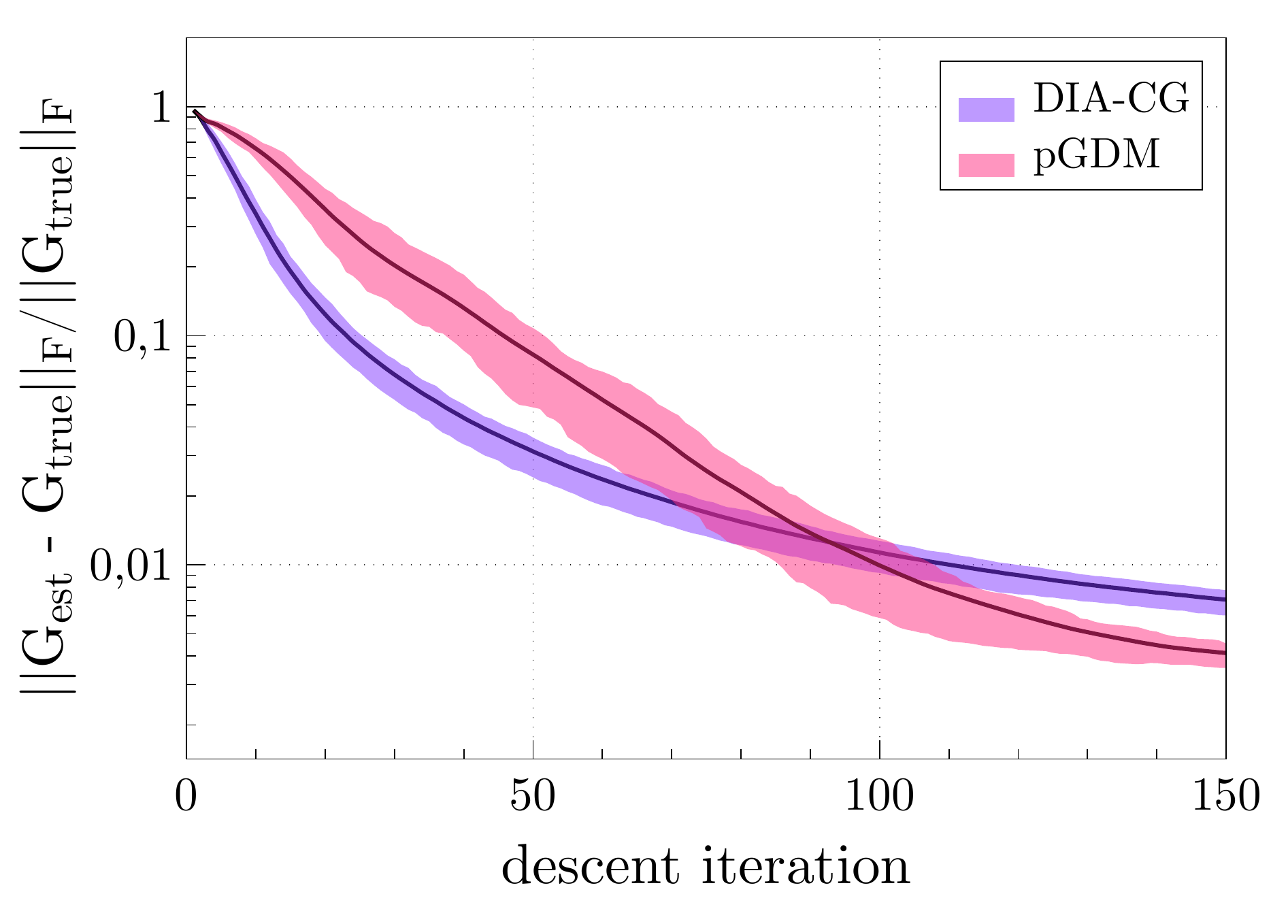}
    \caption{{\bf Comparison of DIA  and pGDM for linearized ML-LQT}. We repeat the linearized ML-LQT for 100 random  $G_{\rm true}$ as in Fig.~\ref{fig:DIA_vs_ALG}, but now considering two different possibilities: (upper panel) well-conditioned matrix $G_{\rm true}$, and (lower panel) ill-conditioned matrix $G_{\rm true}$. We represent the mean and 20/80 percentiles for the estimated $\hat{G}_{\rm lin}$ according to DIA (purple) and  pGDM (pink), as a function of the number of descent iterations. All the definitions are the same as those detailed in the caption of Fig.~\ref{fig:DIA_vs_ALG}.
    }
    \label{fig:dia_vs_pCGD}
\end{figure}

In Fig.~\ref{fig:DIA_vs_ALG}, we show how both the full-CG~\eqref{LT_MLE} (pink) and linear DIA~\eqref{eq:lin_ML} (purple)  minimizations yield a value of the estimated Lindblad matrix $\hat{G}$ that that gets closer to ${G}_{\rm true}$  with each descent iteration. One can see how, on average, the convergence of the linear DIA requires less descent iterations than the full ML-QT, which is further supported by the guaranteed convergence of DIA to the global minimum ensured by the convexity of the linearized estimator. Moreover, the calculation of the gradient for the full ML-LQT requires that, at each step, one must solve the full Lindblad master equation with various values of the estimated matrix $G$, instead of using the simpler linearized expressions for the gradient. Therefore, in addition to the increased number of descent steps in the full ML-LQT, each of them requires more classical computational resources. 
This results in the limited precision of the full nonlinear CG together with a significant runtime slowdown of optimization compared to linear DIA; we observed on average $\approx\!50\times$ and $\approx\!900\times$ difference of optimization time-per-iteration for 1 and 2 qubit estimations respectively, testing on a single-core CPU. As shown in the figure, the full-GC algorithm shows a slower convergence and a wider confidence interval.
On the contrary, the linear DIA shows a much smoother convergence towards the true Lindblad matrix, even after   a small number of  iterations, as it learns any random Lindbladian with a similar accuracy showcasing its higher robustness. After a number of descent steps, the DIA error begins to saturate due to the limitations of the inherent linear approximation. In contrast, the full-CG minimization continues to improve, provided that one allows for a sufficiently large number of descent iterations. As expected, 
 the linearized ML-LQT ~\eqref{eq:lin_ML}  is   a leading-order approximation, and one should not  aim  at accuracies  for which other  higher-order corrections may become dominant.

In summary, the full ML-LQT can reach arbitrary precision in this idealised shot-noise-free situation, at the expense of a much higher-cost in post-processing times and less robustness due to the the lack of convexity. We note that this can be a limitation if the ML-LQT is used as a real-time diagnosis tool to calibrate experimental devices, especially if the noise changes during the post-processing time. In a more realistic situation in which shot noise and other SPAM errors would be present, the accuracy of the Lindbladian reconstruction cannot reach arbitrary precision, as we will see in more detail  below.
In the weak noise limit, the higher-order corrections to the already-small effect of the noise may lead to even smaller contributions that get completely overshadowed by the shot noise, unless one  repeats the experiment  an extremely large number of times, which can be prohibitive in QIP platforms with relatively long cycle times, such as trapped ions. Therefore, in such a situation, aiming at  a higher accuracy by switching to the full ML-LQT  might actually  be counterproductive. 

Once the advantages of the linearized vs full ML-LQT have been identified, we can compare  two different optimization routines for the linearized case. We have so far explored the convergence of the DIA, which deals with the positive-semidefinite constraint of $G$ explicitly by using a Cholesky decomposition. As discussed in App.~\ref{eq:pCGM_CG_app}, an alternative approach based on projected  gradient descent with momentum (pGDM)  allows for updates on the estimated $\hat{G}$ which do not satisfy the constraint, but instead project back to the set of  physically-admissible Lindbladians in a subsequent step. These methods have been proven to be more efficient in the context of QT of states with a high purity (low rank)~\cite{Bolduc2017}, as they lie very close to the boundary of the space of physically-admissible states. In this case,  imposing the positive semidefinite and unit trace constraints at all steps can dramatically slow down the convergence. In the case of linearized ML-LQT, the Lindblad matrix $G$ has no trace constraints as in state-QT, so
it is not a priori clear if similar benefits will be found when running a pGDM algorithm for the estimation of a low-rank Lindblad matrix. In order to make a quantitative comparison of the convergence of the two linearized  ML-LQT algorithms, we also take random samples of low-rank matrices $G_{\rm true}$ by  considering  Haar uniformly random projectors.

Once more, we focus on the $N=2,d=4$ case, using randomly generated two-qubit Lindblad matrices $G_{\rm true}$, and setting their scale so that  
with $(\text{Tr}\{G_{\rm true}\})t_{\rm f} = 0.01$. The comparison for these linear estimators is shown in
Fig.~\ref{fig:dia_vs_pCGD}, where we recall that we are still focusing on the idealised shot-noise-free regime. In the upper panel,
when $G_{\rm true}$ is sampled from the Hilbert-Schmidt uniform set and thus has  many similar eigenvalues,
DIA  (purple) exhibits faster  convergence than the pGDM (pink). However, when $G_{\rm true}$ is a random  projector, we observe in the lower panel that pGDM takes over, converging to closer estimates $\hat{G}_{\rm lin}^{\rm pGDM}$  than the DIA $\hat{G}_{\rm lin}^{\rm DIA}$ for a sufficient number of descent iterations. For a specific range  of descent iterations, the pGDM method shows a 
much quicker descent for low-rank Lindblad matrices than for uniform ones, although we also note that there is a larger variance depending on the specific random Lindbladian one is aiming to learn, as compared to the DIA. In summary, we can conclude that, if one suspects that  an experiment will be affected by  a leading noise source such  that  $G_{\rm true}$
 will only have few substantial eigenvalues, pGDM methods are favoured with respect to DIA. In the following section, we will discuss how the general linearized strategy, regardless of the DIA or pGDM approach used for the gradient descent,  can actually be improved even further in situations where the noise is really structured by using compressed-sensing techniques.

\section{\bf Compressed sensing LQT with structured noise}
\label{sec:04_LLCS}

The number of parameters that must be estimated in  LQT $d^2(d^{2}-1)$ scales exponentially with the qubit number $d=2^N$. As  noted by the end of last section, however, the Lindblad matrix may  have a much  smaller number  of leading jump operators~\eqref{eq:lindblad_equation} that describe the main sources of noise.  As discussed in App.~\ref{app:QT_complexity}, the  operator-sum representations of the  dynamical maps~\eqref{eq:Krauss} associated to such a reduced Lindbladian  would thus have a low Kraus rank,  which can be exploited by compressed sensing (CS)~\cite{kosut2009quantum,PhysRevLett.106.100401,Flammia_2012, PhysRevA.90.012110,Kliesch2019guaranteedrecovery}. In fact, for   process-QT~\cite{PhysRevLett.106.100401,Rodionov2014}, when one knows the specific unitary operator that lies close to the actual time evolution, compressed sensing can be optimised to reduce the number of configurations to a  polynomial one by exploiting the sparseness of the process matrix in a certain basis.  To  our knowledge, the application of  compressed-sensing techniques to  ML-LQT has not been considered yet. In this section, we explore how this prior  on the noise structure  can also be exploited for our linearized ML-LQT,   allowing us to obtain a faithful estimation of the  Lindblad matrix  with a much lower number of measurements.

\begin{figure*}[t]
\centering
\includegraphics[width=\linewidth]{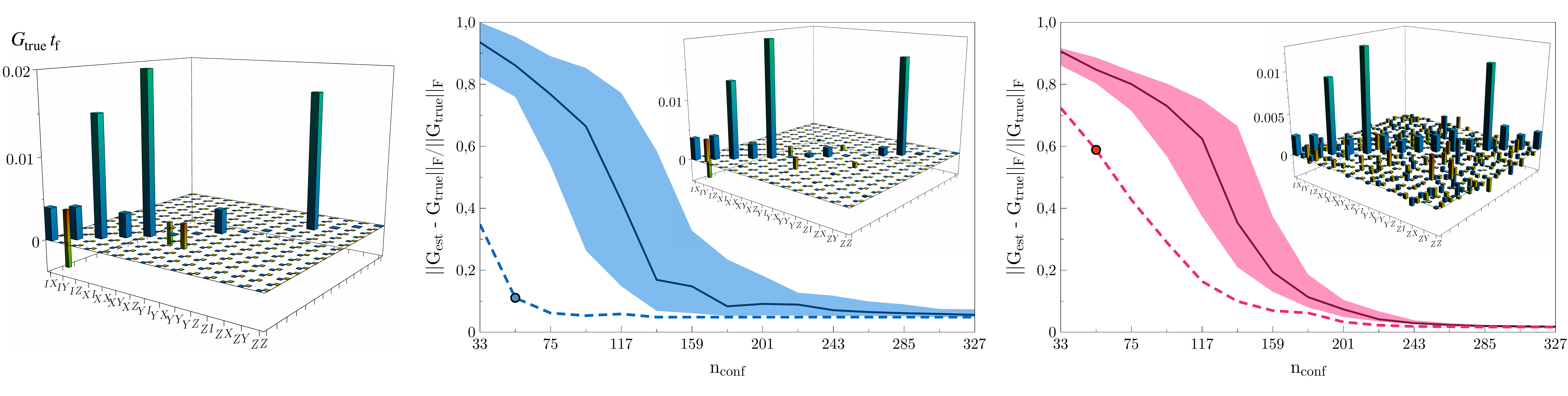}
\caption{ {\bf Comparison of compressed sensing and full ML-LQT:} We follow the same prescription described in the caption of Fig.~\ref{fig:DIA_vs_ALG}, but now considering a single true $G_{\rm true}$, a sufficient number of descent iterations for any of the CS or full ML-LQT estimation methods, and plotting the relative Frobenius norm of the estimation error as a function of the number of configurations. (Left panel) Skyline plot of the decomposition of the Lindblad matrix $G_{\rm true}$ in the Pauli basis. 
(Middle and right panel) Frobenius norm distance of  $\hat{G}_{\rm true}$ and $\hat{G}_{\rm lin}$  estimated with either CS-LQT with $\epsilon=1.2 $ (blue) or full  ML-LQT (pink). We consider again a single evolution time $t_i=t_{\rm f}$, and iteratively grow the number of  configurations by increasing the different initial states and measurement POVMs that are included in the corresponding cost functions/constraints. In particular,  we grow randomly from initial batches with $n_{\textrm{conf},i}=33$ by adding $\delta n_{\textrm{conf},i}=21$ at each step. The random selection of the configurations is repeated $50$ times, and we present the median and the 20/80 percentiles of the results. 
For $n_{\textrm{conf},i}\geq240$ (QPT informational completeness) both methods attain a similar reconstruction of $G_{\textrm{true}}$. On the other hand, for a number of configurations well below that number, the CS estimation shows a better accuracy and a higher dispersion of the data, as some of the random configurations may give less information for the CS estimations. The dashed lines represent the optimal sequence of configurations for both methods among the 50 repetitions, highlighting  that  CS can attain a very high accuracy already for low   configurations $n_{\rm conf,i}=54$ (circles). The corresponding estimated Lindblad matrices are displayed in the insets which, by direct comparison to $G_{\rm true}$ on the left panel, showcase the improvement of CS-LQT with respect to ML-LQT.}
\label{fig:LCS_2q_testbed}
\end{figure*}

CS originates in the classical theory of signal processing, where the signal is described by a certain sparse vector that can be  recovered from an under-sampled set of measurements 
~\cite{Donoho2006, Candes2006}.  A standard 
approach to maximise  sparsity is to minimize the $1$-norm of the signal vector, defined as the sum of the absolute values of its components.
The incorporation of  CS to  process-QT allows for the estimation of the process matrix using informationally incomplete data sets 
by defining a CS estimator that resembles this norm  applied to the process matrix~\cite{Kosut2008}. 
In the present work, we  use such a CS estimator to learn the Lindblad matrix in the linear regime of Eq. \eqref{eq:est_prob_lin}. 
This  procedure maximises the sparsity of the Lindblad matrix subject to a constraint with which   the Lindblad model  must be able to reproduce the  observed measurement outcomes with a certain error $\epsilon$ with respect to a least-square measure of the distance between the theoretical and measured probability distributions. This can be formalised through the following constrained convex minimization problem 
\beq
\label{CS_LT}
\begin{split}
\hat{G}_{\rm lin}^{\rm CS}=&\texttt{argmin}\big\{\mathsf{C}_{\rm CS}(G)= \sum_{\alpha,\beta}\!\big(|\text{Re}(G_{\alpha\beta})| + |\text{Im}(G_{\alpha\beta})|\big)\big\}\\
&\texttt{subject to}\; \, \big\|\, \boldsymbol{f} - \boldsymbol{p}^{\rm u}  - \sum_{\alpha,\beta}G_{\alpha\beta}\boldsymbol{\Phi}^{\alpha\beta}
\,\big\|_{2}\!\! < \sqrt{n_{\textrm{conf}}}\,\varepsilon.
\end{split}
\eeq
Here, $\boldsymbol{f}, \boldsymbol{p}^{\rm u} \in \mathbb{R}^{n_{\textrm{conf}}}$ are, respectively,  the observed relative frequencies and the contribution of the ideal unitary evolution to the estimated probabilities~\eqref{eq:p_U},  both of which have been vectorised   in the space of configurations. The total number of configurations per time step is  $n_{\textrm{conf},i} =3^Nd^2(d-1)$, although we remark that the CS method  will  use a smaller number as we are interested in reducing the sampling complexity. Likewise, we have introduced a  set of vectors $ \boldsymbol{\Phi}^{\alpha\beta} \in \mathbb{C}^{n_{\textrm{conf}}}$ by reshaping the linearized contribution of the  Markovian dynamics~\eqref{eq:lin_contrib_noise} to the estimated probabilities~\eqref{eq:est_prob_lin}. Let us remark that the distance between the measured and estimated probabilities is no longer present in the cost function, but instead appears in the constraint. While the previous ML estimation aims at inferring the Lindblad model with which one could predict the observed outcomes with a higher likelihood,  CS attempts to find the  sparsest Lindblad model at the expense of  obtaining  a lower  probability to reconstruct the observations. In fact, the above $\epsilon$   acts as a trade-off parameter, as it  balances the tendency to fit the  experimental data minimizing the least-square distance or maximizing  the sparseness of  $\hat{G}$.

Although, as advanced in the introduction and explained further in App.~\ref{app:QT_complexity},  CS  for process-QT can get considerable improvements even without prior knowledge of the sparsifying basis~\cite{Flammia_2012,Kliesch2019guaranteedrecovery}, one can get further improvements towards a polynomial scaling when this is known~\cite{PhysRevLett.106.100401,Rodionov2014}. This basis dependence becomes very transparent at the level of the Markovian jump operators~\eqref{eq:lindblad_equation}. To quantify these possible improvements, we start by 
 considering  LQT for  a 2-qubit system as before, but    considering a restricted set of  Markovian noise  jump operators $L_n\in\{L_{\rm deph,1},L_{\rm deph,2},L_{\rm damp,1},L_{\rm damp,2},L_{\rm bf}\}$ instead of randomly sampling an unstructured  Lindblad matrix. We include  single-qubit dephasing with  rates $\gamma_{\rm deph,1} \ne \gamma_{\rm deph,2}$ and jump operators
  $L_{\rm deph,1} = \sigma_z\otimes\mathbb{1}_2$, $L_{\rm deph,2}  = \mathbb{1}_2\otimes\sigma_z$,  single-qubit amplitude damping with rates  $\gamma_{\rm damp,1} \ne \gamma_{\rm damp,2}$ and  operators
  $L_{\rm damp,1} = \sigma_-\otimes\mathbb{1}_2$, $L_{\rm damp,2} = \mathbb{1}_2\otimes\sigma_-$ and, finally, 
 correlated bit-flip errors  $L_{\rm bf}=\sigma_x\otimes\sigma_x$ with a rate $\gamma_{\rm bf}$.  All these decay rates and associated jump operators will define the true Lindblad matrix  $G_{\rm true}$ we aim at learning.

In light of these jump operators and in order to exploit the sparseness of the Lindbladian, we use the Pauli basis~\eqref{eq:pauli_basis} in the linearization.
To obtain an estimate $\hat{G}_{\rm lin}^{\rm CS}$ via Eq.~\eqref{CS_LT}, we need to calculate the linearized contributions to the predicted probabilities~\eqref{eq:est_prob_lin}. In the present case, we set $H=0$ such that the ideal  evolution is the identity and we are simply estimating the qubit decoherence as a quantum memory. Hence,  the vector $\boldsymbol{p}^{\rm u}$~\eqref{eq:p_U} can be readily obtained.  Finally, the set of vectors $ \boldsymbol{\Phi}^{\alpha\beta}$~\eqref{eq:lin_contrib_noise}  require a numerical evaluation of Eq.~\eqref{eq:delta_rho_lin} which, ultimately, requires using $b_{\alpha}^p=\delta_{p,\alpha}$ in Eqs.~\eqref{B_tensor}-\eqref{c_tensor} for the Pauli-basis choice. Once all these quantities are at our disposal, we  employ the general augmented Lagrangian algorithm from the C++ ALGLIB library~\cite{ALGLIBweb} for the constrained minimization~\eqref{CS_LT} choosing a target  $\epsilon$ parameter.

The main motivation for turning to CS-LQT techniques is that one can obtain accurate estimations using a smaller set of configurations. We recall that, for the 2-qubit case, the total number of configurations in the ML-LQT is $n_{{\rm conf},i}=3^Nd^2(d-1)=432$ per time step, while informational completeness   requires  $d^2(d^2-1)^2=240$ linearly-independent configurations.  In order to address the accuracy of the CS-LQT in under-sampled situations, we randomly draw a small number of triples $(\rho_{s,0},t_i,M_{\mu})$ from all possible configurations, and gradually increase the set by subsequently including $\delta n_{{\rm config}, i}$ additional  random configurations until all of the $n_{{\rm conf},i}=432$ configurations are incorporated, considering again a single $|\mathbb{I}_t|=1$ snapshot at $t=t_{\rm f}$. In this informationally-complete regime,  the  CS-LQT should give similar estimates to the linearized ML-LQT.

For each CS estimation~\eqref{CS_LT}, which uses a specific choice of configurations for which the measurement data would be collected, we calculate the Frobenius  norm of the difference between the estimated $\hat{G}_{\rm lin}^{\rm CS}$ and  the true sparse Lindblad matrix $G_{\rm true}$, which is depicted in the left panel of  Fig.~\ref{fig:LCS_2q_testbed}. The central and right panels illustrate the result of our numerical simulations, where the true $G_{\rm true}$ is used to generate the $f_{s,i,\mu}\mapsto p_{s,i,\mu}(G_{\rm true})$ probability distribution, dispensing again with the effects of shot noise and SPAM errors for the moment. We depict the normalised Frobenius distance between the estimated and true Lindblad matrices as a function of the number of configurations being included in the linear CS-LQT (central panel) and ML-LQT (right panel) estimators. The solid lines represent the mean, while the shaded areas are the 20/80 confidence intervals, showing how  the accuracy of the estimation grows for different random  ways in which the number of configurations is increased. In addition, the dashed lines represent the optimal configuration choice for this specific $G_{\rm true}$ among the 50 repetitions, with a filled circle that indicates the Frobenious distance for $n_{{\rm config},i}=54$, being the corresponding estimated $\hat{G}_{{\rm lin}^{\rm CS}},\hat{G}_{{\rm lin}^{\rm DIA}}$ represented in the corresponding insets. The comparison of these dashed lines shows that the  CS-LQT strategy can reach a much more accurate estimate for an under-sampled data set than the linear ML-LQT approach. As we can directly see in the insets, the CS-LQT estimated Lindblad matrix resembles more closely the true one (left panel), whereas the linear ML-LQT estimation is corrupted by many small non-zero coefficients, lowering considerably the overall estimation accuracy. The superiority of CS-LQT is a consequence of the sparseness of the true Lindblad matrix $G_{\rm true}$, and the underlying sequence of configurations underlying the optimal CS-LQT estimation are those that providing more relevant information about the noise, determined by the Lindblad matrix eigenbasis. The identification of this eigenbasis can be based on microscopic knowledge about the QIP, or from the knowledge acquired by running full ML-LQT for small system sizes, and then extrapolating the conclusions about the noise structure to larger systems.

In summary, 
ML-LQT  chooses to retain the small components of the estimated Lindblad matrix in order to reach a higher likelihood between the measured and   estimated probabilities. However, this  leads to a larger error with respect to CS-LQT in undersampled regimes, as the latter neglects almost all of those very small components by minimising the sparseness-based cost function~\eqref{CS_LT}.
 The CS estimate  captures the most prominent noise components at this level, providing  more than  a ten-fold reduction in the number of shots with respect to the ML-LQT with all 432 configurations. Therefore, if a large number of  measurements are inaccessible experimentally, or one scales to larger system sizes where the shots need to be distributed among more configurations, CS-LQT will offer a superior solution to ML-LQT if one has some prior knowledge about the structure and sparseness  of the noise.

\section{\bf Single- and two-qubit    trapped-ion LQT } 
\label{sec:06_ions}

Trapped-ion systems have played a key role in the development of quantum computers.
In the seminal work~\cite{PhysRevLett.74.4091}, it was proposed that ion crystals can function as registers for the realization of a quantum computer. In this setup, quantum information is encoded in the electronic levels of the ions and manipulated using a universal gate set that involves additional lasers to excite their collective vibrations. This proposal, which was first implemented in~\cite{Schmidt-Kaler2003}, paved the way for extensive experimental and theoretical work, establishing trapped ions as one of the leading platforms in the pursuit of constructing fault-tolerant quantum computers~\cite{HAFFNER2008155,10.1063/1.5088164}.

At present, trapped-ion QIPs have served to experimentally realize various noisy intermediate-scale quantum (NISQ) algorithms  over the years~\cite{Gulde2003,Barrett2004,Riebe2004,doi:10.1126/science.1110335,doi:10.1126/science.aad9480,Figgatt2017,doi:10.1126/science.aaw9415}. Moreover, there are  ongoing  efforts in developing trapped-ion quantum error correction~\cite{Chiaverini2004,Schindler1059,Nigg302,Linkee1701074,Negnevitsky2018,Flühmann2019,Stricker2020, https://doi.org/10.48550/arxiv.2010.09681,Erhard2021, Egan2021, PhysRevLett.127.240501,PhysRevX.11.041058,PhysRevX.12.011032,Postler2022,https://doi.org/10.48550/arxiv.2208.01863}, and developing a detailed microscopic noise modelling to assess  the quantum error correction (QEC) performance   under   realistic experimental conditions~\cite{PhysRevX.7.041061,PhysRevA.99.022330,PhysRevA.100.062307,ParradoRodriguez2021crosstalk,rodriguezblanco2022witnessing}. The success of NISQ and QEC endeavours heavily relies on the high-fidelity universal gate set native to the trapped-ion platform~\cite{fidelities1,PhysRevLett.117.060504,PhysRevLett.117.060505,PhysRevLett.117.140501,Erhard2019,PhysRevLett.123.260503}.
Trapped ions have also pioneered several QT experiments for the characterization of    entangled states~\cite{Leibfried1996,PhysRevLett.92.220402,Lanyon2017,Häffner2005,PRXQuantum.3.040310}, as well as single- and multi-qubit gates~\cite{PhysRevLett.97.220407,doi:10.1126/science.1177077, Riebe_2007, PhysRevLett.102.040501,PhysRevLett.103.200503,PhysRevA.81.062332,PhysRevA.90.010103,Tinkey_2021}. In fact, many pioneering  schemes for QT, such as randomised benchmarking~\cite{Emerson_2005,PhysRevLett.106.180504} and extensions thereof~\cite{PhysRevA.94.052325}, or gate set tomography~\cite{blumekohout2013robust}, were first implemented with trapped ions~\cite{PhysRevA.77.012307,PhysRevA.84.030303,PhysRevLett.108.260503,PhysRevLett.117.060504,Blume-Kohout2017,Mavadia2018, Erhard2019,Wright2019,Pogorelov2021}. In the  context of LQT,  there has only been one previous experiment to the best of our knowledge, which focused on the spontaneous emission of a single trapped-ion qubit under different engineered decay channels~\cite{PhysRevA.101.062305}.  However,  LQT has not been applied to learn the Lindblad generators of  real non-injected noise in high-fidelity gates including, in particular, LQT for  noisy two-qubit gates. The goal of this section is to fill in this gap using data  from the experiments in~\cite{Schindler2013, Schindler2013b,Pogorelov2021}.

In order to apply the linearized ML and CS tools for LQT in real trapped-ion systems, we first need to reconsider the above strategies for a  finite number of measurements $N_{\rm shots}$, such that $ p_{s,i,\mu}\approx f_{s,i,\mu}$. As discussed below Eq.~\eqref{eq:log_likelihood_Lindblad} and in Appendix~\ref{app:QT_complexity}, one records  the number of times $N_{s,i,\boldsymbol{b},\boldsymbol{m}_{\boldsymbol{b}}}$ that the  $\boldsymbol{m}_{\boldsymbol{b}}$-outcome is observed   for each     initial state $\rho_{0,s}$,  evolution   time $t_i\in T$, and measurement basis $\boldsymbol{b}$. We consider that  $N_{\rm shots}$ are equally distributed among each  setting  $N_{\rm shots}=3^Nd^2|\mathbb{I}_t|\times N_{\rm sc}$,  such that $N_{\rm sc}=\sum_{\boldsymbol{m}_{\boldsymbol{b}}}N_{s,i,\boldsymbol{b},\boldsymbol{m}_{\boldsymbol{b}}}$ is the same   $\forall s,i,\boldsymbol{b}$, and we can obtain the  relative frequency by simply taking a ratio $f_{s,i,\boldsymbol{b},\boldsymbol{m}_{\boldsymbol{b}}}=N_{s,i,\boldsymbol{b},\boldsymbol{m}_{\boldsymbol{b}}}/N_{\rm sc}$. The  ML-LQT and CS-LQT strategies presented in the previous sections will not only be limited by the effects discussed previously, such as the accuracy of the non-linear minimization or the number of configurations included in the estimator, but also by stochastic errors associated to this shot noise. In addition, SPAM errors  in the state-preparation and measurement  will also affect the inference.

Another  point that has not  been addressed in detail yet is that, when learning from real data, we do not know the true Lindblad matrix, and thus cannot provide estimates for the accuracy of  our estimates, or  find how many descent iterations and how many configurations would be required  to reach an specific  target. We  now describe a simple way in which  the readout data is not only used  to extract the model parameters by ML or CS, but also for  error analysis and hypothesis testing. In principle, if one could collect large numbers of data, it would be possible to use the multinomial distribution of the likelihood function to perform a statistical analysis of errors and confidence intervals for the estimation~\cite{velazquez2024dynamical}. In this section, we follow a much simpler strategy that could actually be performed during the gradient descent or as one increases the number of configurations, providing us with a simple criterion of convergence: the  Pearson $\chi_{\rm P}^2$ test~\cite{Taylor1982}. This test assesses the goodness of a fit between a joint multinomial distribution with probabilities parametrized by the estimated Lindbladian $\hat{\boldsymbol{c}},\hat{G}$, and the finite-frequency approximation to the  distribution measured in the experiment. The reduced Pearson  $\chi^2_{P}$  statistics reads
\begin{equation}
\chi^2_{\rm P} = \frac{N_{\rm shots}}{(d-1)} \sum_{s,i,\boldsymbol{b},\boldsymbol{m}_{\boldsymbol{b}}}\!\! \frac{(p_{s,i,\boldsymbol{b},\boldsymbol{m}_{\boldsymbol{b}}}(\hat{\boldsymbol{c}},\hat{G}) - f_{s,i,\boldsymbol{b},\boldsymbol{m}_{\boldsymbol{b}}})^2}{p_{s,i,\boldsymbol{b},\boldsymbol{m}_{\boldsymbol{b}}}(\hat{\boldsymbol{c}},\hat{G}) }.
\label{eq:chi2_pears}
\end{equation}
This quantity converges to the reduced $\chi^2$ of a Gaussian distribution with variance $\sigma_{\chi}^2=2/(d-1)$  in the large-$N_{\rm shots}$ limit, and can be used as a simple indicator of goodness of fit:  $\chi_{\rm P}^2 \gg 1$ when
the parametrized probabilities are not an adequate model to reproduce the data, and $\chi_{\rm P}^2 \ll 1$ when
there is overfitting, and  the theoretically parametrized probabilities  are improperly accounting for the noise, or the underlying error bars would be over-estimated. A heuristic converge criterion is to stop when the Pearson $\chi^2_{\rm P}$ falls below unity $\chi^2_{{\rm P,c}}=1$.
An important note on the  applicability of the method
is that the number of counts in any of the ``bins''
$N_{\rm sc}f_{s,i,\boldsymbol{b},\boldsymbol{m}_{\boldsymbol{b}}}$ should not be too small~\cite{Taylor1982}, which might be the
case for high-fidelity gates. We can now obtain estimates for the  discrepancy between observed outcomes and the values expected under the Markovian Lindblad model as one performs subsequent descents in the non-linear minimizations, or as one includes more configurations in the estimators towards  informational completeness. In this way, we  can decide when to stop the LQT learning without knowing the true Lindbladian.

\subsubsection{LQT for  single-qubit gates}
Let us now consider these aspects for the LQT of trapped-ion quantum gates. We start from the simple single-qubit rotations,
focusing on a $\pi/2$ rotation around the $x$ axis
\begin{equation}
  R_{X}(\Theta(t_{\rm f},t_0)) = \ee^{-\ii\frac{\Theta(t_{\rm f},t_0)}{2}\sigma_x}, \hspace{1ex}\Theta(t,t_0)=\!\int_{t_0}^{t_{\rm f}}\!\!\!{\rm dt'}\Omega(t'),
\end{equation}
  where $t_{\rm f}-t_0$ is the gate time, and $\Omega(t')$ defines a possibly-modulated Rabi frequency for the specific qubit transition. The ideal target unitary $U(t_{\rm f},t_0)=(\mathbb{1}_2-\ii\sigma_x)/\sqrt{2}$ is obtained by setting $\Theta(t_{\rm f},t_0)=\pi/2$ in the  expression above. As trapped-ion QIPs routinely achieve very high fidelities for single-qubit gates~\cite{10.1063/1.5088164}, we can start by assuming a perfect knowledge of this unitary, and focusing the LQT on the estimation of the unknown Lindbladian matrix $G_{\rm true}$. We use
the measurement data from the $^{40}{\rm 
Ca}^+$ experiments by P. Schindler et al. ~\cite{Schindler2013, Schindler2013b},
in which  the configurations include $|\mathbb{S}_0|=4$ initial states, a single  time at the end of the gate $|\mathbb{I}_t|=1$, and $|\mathbb{M}_{b}|=3$ for the measurements in the $b\in\{x,y,z\}$ single-qubit basis, each of which has a  binary Pauli outcome $m_b\in\{-1,+1\}$. Therefore, the  total number of independent configurations is $n_{\rm conf}=12$, among which the  total number of measurements $N_{\rm shot}=1.2\cdot 10^5$  were distributed uniformly. Note that the number of configurations equals the required number of Lindbladian parameters $d^2(d^2-1)=12$ if one were to  learn both the Hamiltonian and the Lindbald matrix. To gauge the relevance of  the shot noise, we start by running our previous LQT algorithms on numerically simulated data, where we sample $N_{\rm shots}$ times from the ideal probability distribution $
p^{\rm u}_{s,i,b,\sigma}$~\eqref{eq:p_U}. We thus generate $N_{\rm shots}$ uniform random numbers and numerically simulate the Bernoulli trials, collecting the number of obtained outcomes $N_{s,i,b,\sigma}$  for a fixed number of  shots $N_{\rm sc}=10^4$  per initial state, evolution time and measurement basis. In the present case, we have a single evolution time $t_i=t_{\rm f}$. In this way, we simulate the effects of shot noise in the finite frequencies $f^{\rm u}_{s,i,b,\sigma}$. As there are no other sources of noise, one expects that running the LQT algorithms will provide a structureless Lindblad matrix $\hat{G}_{\rm u}$, leading to estimated jump operators $\{\hat{L}_n^{\rm u}\}_{n=1}^3$~\eqref{eq:lindblad_equation} that have no preferred direction in the qubit's Bloch sphere. On the other hand, the eigenvalues $\gamma^{\rm sn}_n$  can be used to quantify the level  of the contribution of the shot noise to the decay rates
$\overline{\gamma}_{{\rm sn}}={\rm max}\{\hat{\gamma}_{n}^{{\rm u}}: n\in\{1,2,3\}\}$. For $N_{\rm shots}=1.2\cdot 10^5$ shots, we obtain  $\overline{\gamma}_{{\rm sn}}\approx 0.8\cdot10^{-3}$. This sets a lower bound below which the estimated decay rates, which will be derived  using  the real experimental data below,  would be dominated by the  shot noise and thus be inconclusive.

\begin{figure}[!t]
  \centering
  \includegraphics[width=\linewidth]{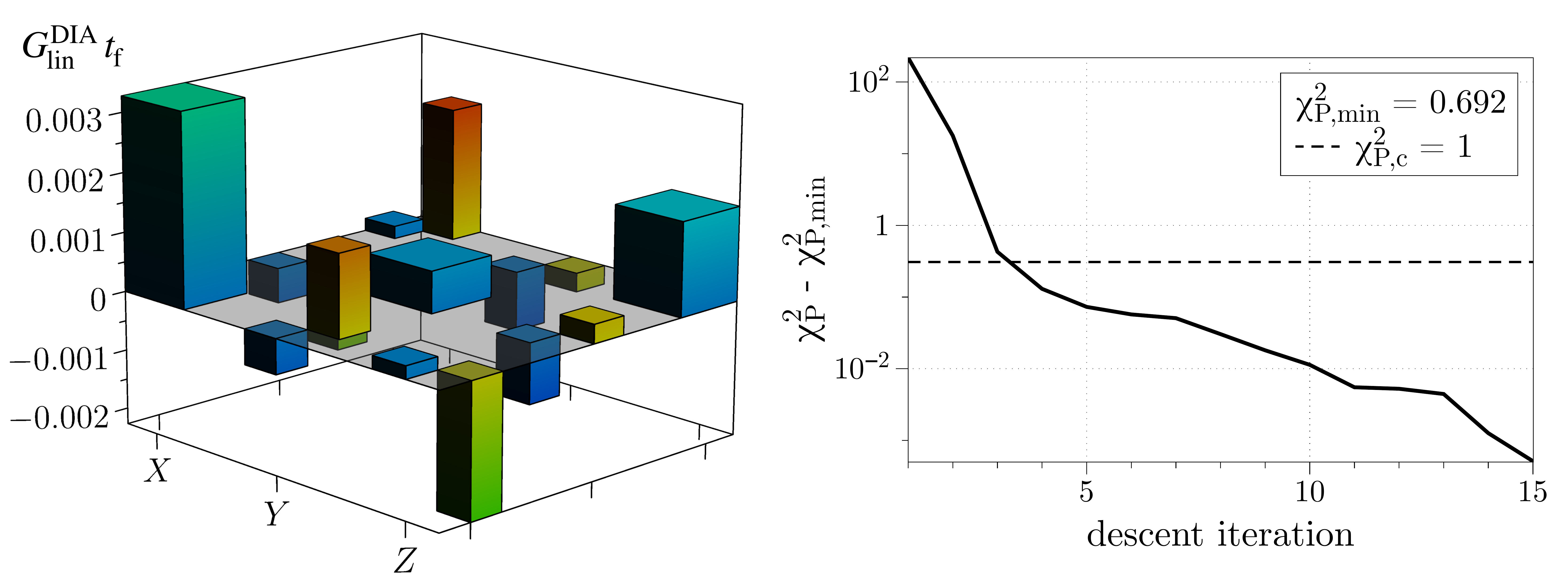}
  \caption{{\bf linearized ML-LQT for  trapped-ion single-qubit gates:} We consider the informationally complete setting with $n_{\rm config,i}=12$ configurations, each measured projectively with $N_{s,i,b}=10^4$ shots, and solve the linearized ML-LQT~\eqref{eq:lin_ML} with our DIA. (Left panel) Skyline plot of the estimated Lindblad matrix $\hat{G}^{\rm DIA}_{\rm lin}$, showing predominant contributions for elements involving the $X$ and $Z$ Pauli matrices. (Right panel) Pearson $\chi^2_{\rm P}$ as a function of the gradient descent iteration. After just a few interactions, starting from a diagonal $G$ with $\rm{Tr}\{G\}t_{\rm{f}} = 0.25$, we already obtain a value well below the $\chi^2_{\rm P,c}=1 $, signalling a reasonable convergence.}
  \label{fig:Rx_gate_full_ ML-LQT}
\end{figure}

Let us now analyse the real experimental data which, in addition to shot noise, will also be afflicted by SPAM errors. In the left panel of Fig.~\ref{fig:Rx_gate_full_ ML-LQT}, we display the results of the estimated Lindblad matrix $\hat{G}_{\rm lin}^{\rm DIA}$ obtained by solving the linearized ML-LQT~\eqref{eq:lin_ML} for the $\pi/2$ pulse using the DIA for gradient descent. On the right panel, we display the Pearson reduced  $\chi^2_{\rm P}$-test, showing that we fall  below the critical  $\chi^2_{\rm P,c}=1$ very fast as the  number of  descent steps is increased, reaching a minimal value of  $\chi^2_{\rm P,min}=0.695$ that supports a reasonably good fit. By diagonalizing the estimated Lindblad matrix $\hat{G}_{\rm lin}^{\rm DIA}$, we obtain  a single leading decay rate $\hat{\gamma}_{\rm max}=5.6\cdot10^{-3}>\overline{\gamma}_{{\rm sn}}$, while the other two are below the $10^{-12}$ level. Considering that this leading decay rate is almost  10$\times$ bigger than the estimated shot noise,  we can conclude that our DIA ML-LQT is not limited  by shot noise for the number of measurement  $N_{\rm shots}=1.2\cdot 10^5$ performed in the experiment. 
We can thus discern the underlying structure of the sources of noise from the LQT. 
In fact, the jump operator corresponding to the maximum decay rate is found to have the following
decomposition $L_{\rm max} = n_x\sigma_x + n_y\sigma_y + n_z \sigma_z$ with $|n_x| = 0.77$, $|n_y| = 0.36$, $|n_z| = 0.53$, which shows that the main error generator of single-qubit gates is  biased towards the gate rotation axis. 

We note that, for the single-qubit case, we have also run the CS-LQT~\eqref{CS_LT}, but the estimates are similar to those of linear ML-LQT, as the size of the Lindblad matrix is already small and there is no big sparsity that can be exploited. We will show below that CS becomes more useful for two qubits.

\subsubsection{LQT for two-qubit Mølmer–Sørensen gates}

Let us now turn to the LQT of entangling trapped-ion  gates.
In particular, we focus on the two-qubit  Mølmer–Sørensen (MS) gate~\cite{PhysRevLett.82.1971,PhysRevA.62.022311}. This gate exploits the quantized vibrations of the ion crystal, i.e. phonons,  to  generate entanglement between  a pair of   qubits that belong to the same ion crystal. This gate has become a workhorse in trapped-ion quantum computing, as it allows to achieve high-fidelity gates even in the presence of thermal fluctuations, i.e.~without requiring perfectly groundstate-cooled vibrational modes. For two qubits, this unitary gate can be interpreted as the coherent evolution under an Ising-type Hamiltonian 
\begin{equation}
R_{XX}(\Theta(t_{\rm f},t_0))= \ee^{-\ii \frac{\Theta(t_{\rm f},t_0)}{2}\sigma_{x}\otimes\sigma_x}.
  \label{eq:XX_MS}
\end{equation}
Here,  $\Theta(t_{\rm f},t_0)$ is the pulse area that depends on the gate time $t-t_0$,  laser parameters such as intensity and frequency, as well as the frequencies and Lamb-Dicke parameters associated to the vibrational modes that mediate the interactions~\cite{Cai2023}. In particular, assuming that the MS gate is obtained by a pair of  beams of opposite detuning $\pm \mu_L$  with respect to the  qubit transitions,  driving  the sidebands for  the longitudinal center-of-mass mode of frequency  $\nu_z$  near-resonantly for a two-ion crystal, one finds 
\beq
\Theta(t_{\rm f},t_0)=\frac{\eta_{\rm LD}^2\nu_z}{8\pi^2}\frac{\mu_L-\nu_z}{\mu_L+\nu_z}\!\int_{t_0}^{t_{\rm f}}\!\!\!{\rm dt_1'}\!\!\int_{t_0}^{t_{\rm f}}\!\!\!{\rm dt_2'}\Omega(t_1')\Omega(t_2'),
\eeq
where $\eta_{\rm LD}$ is the Lamb-Dicke parameter, and  $\Omega(t)$ is the Rabi frequency. Here, we have assumed that the gate time fulfills $t_{\rm f}-t_0=2\pi/(\mu_L-\nu_z)$, i.e.~a single-loop MS gate is realized, and that the laser intensities are calibrated such that   $\Theta(t_{\rm f},t_0)=\pi/2$, and the target MS gate is $U(t_{\rm f},t_0)=(\mathbb{1}_4-\ii\sigma_x\otimes\sigma_x)/\sqrt{2}$.

We  use the experimental  data gathered on the  $^{40}{\rm 
Ca}^+$ setup described in~\cite{Pogorelov2021}, which includes $|\mathbb{S}_0|=16$ initial states, a single  time at the end of the gate $|\mathbb{I}_t|=1$, and $|\mathbb{M}_{\boldsymbol{b}}|=9$ for the measurements in the two-qubit basis $\boldsymbol{b}\in\{xx,xy,xz,\cdots,zz\}$, each  leading to $|\mathbb{M}_{\boldsymbol{m}_{\boldsymbol{b}}}|=3$ independent outcomes, e.g. $\boldsymbol{m}_{\boldsymbol{b}}\in\{(+1,+1),(+1,-1), (-1,+1)\}$. Therefore, the  total number of independent configurations is $n_{\rm conf}=432$, for which the data set contains a total number of measurement outcomes $N_{\rm shots}=1.44\cdot 10^5$  this time. 
In contrast to the single-qubit case, 
the configuration set now exceeds the $d^2(d^2-1)=240$ real parameters that are required to determine the Lindbladian, as we already discussed in the previous numerical simulations.

\begin{figure}[!t]
  \centering
  \includegraphics[width=0.9\linewidth]{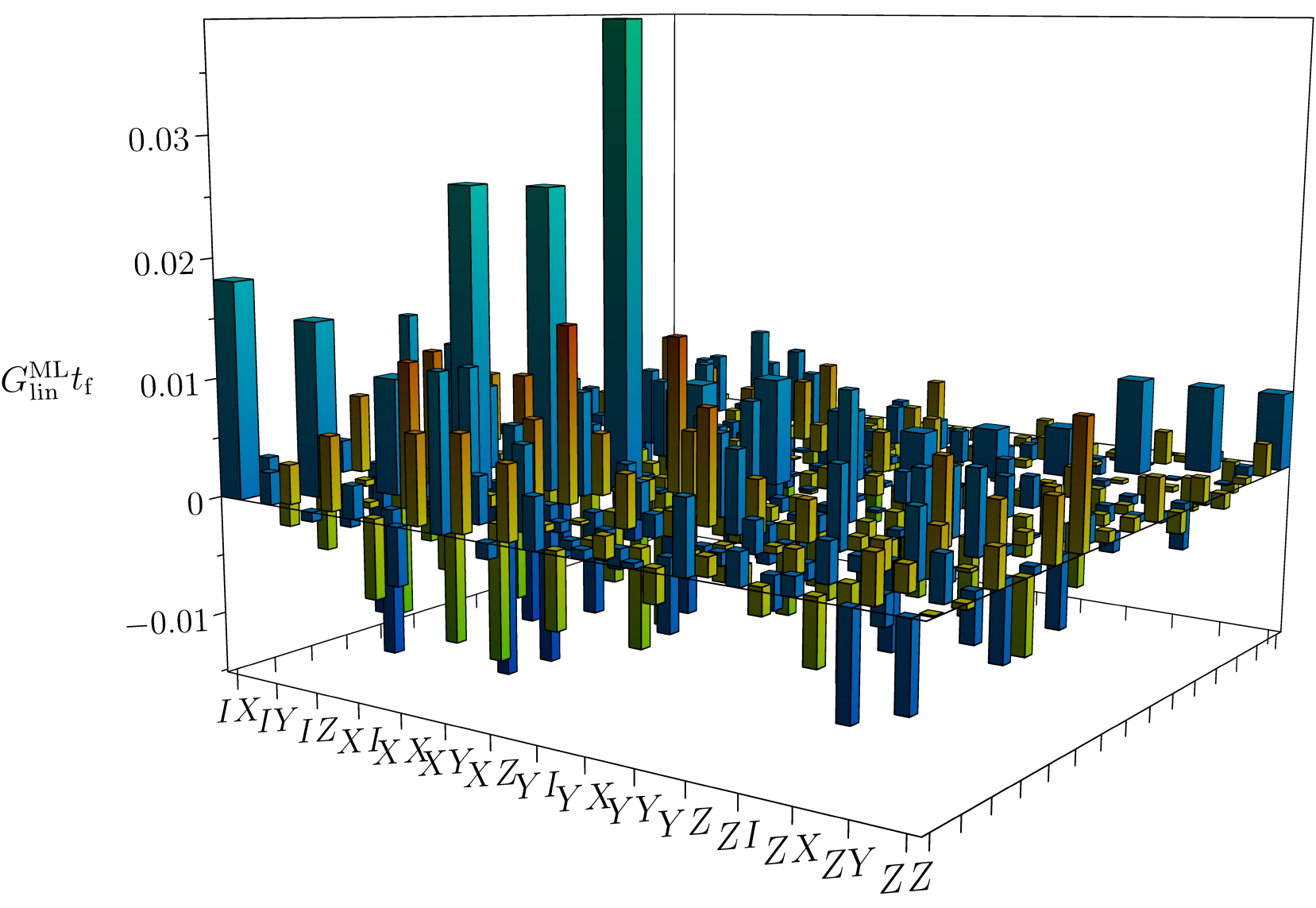}
  \includegraphics[width=0.9\linewidth]{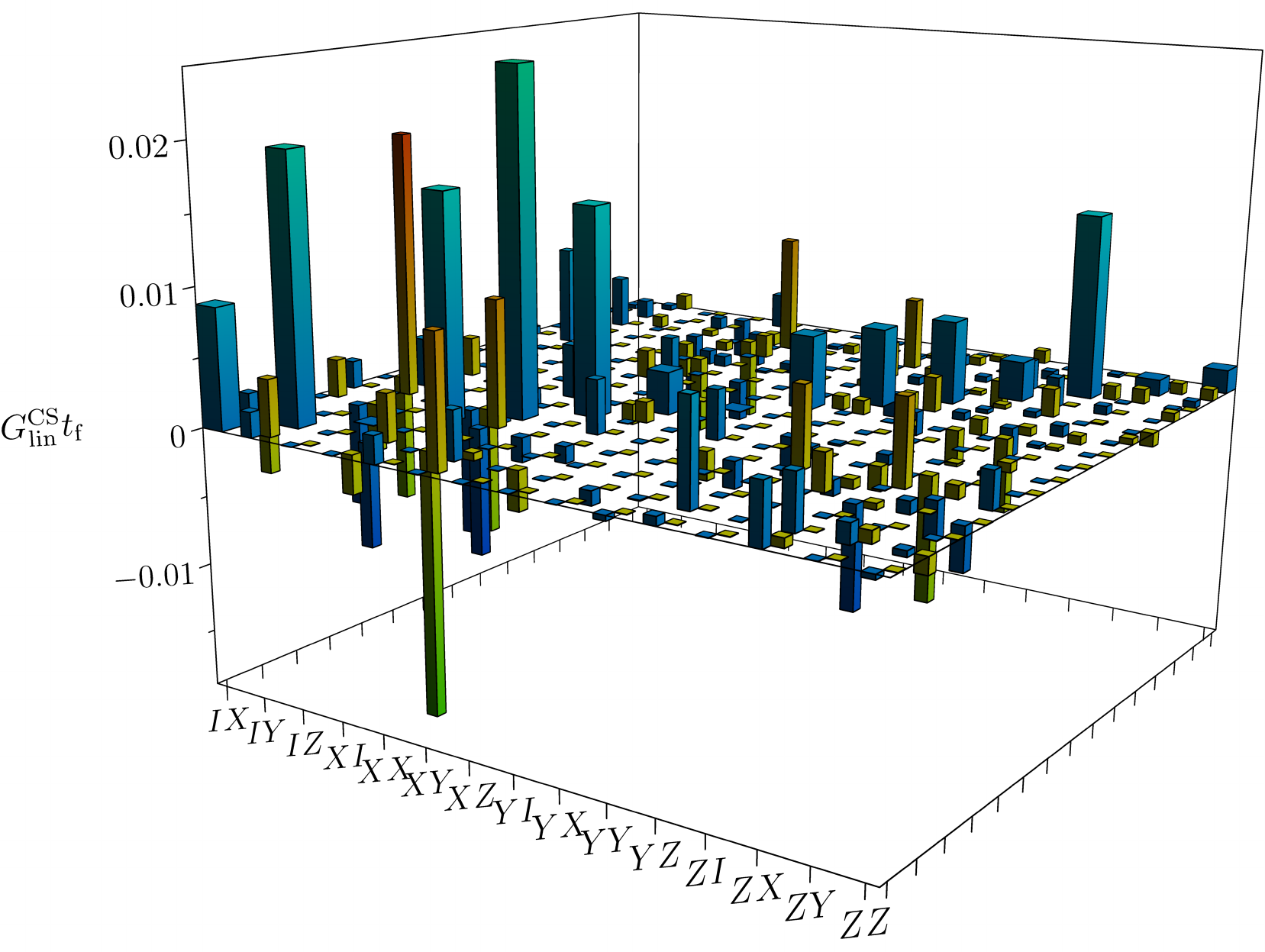}
  \caption{{\bf LQT estimates of trapped-ion two-qubit gates:} Skyline plots of $\hat{G}_{\textrm{lin}}^{\rm DIA}$ obtained using linear ML-LQT (upper panel) and $\hat{G}_{\textrm{lin}}^{\rm CS}$ using CS-LQT  (lower panel), both with with $n_{{\rm config},i}=96$. The  shot-noise has a rough contribution of $0.01$ to the estimated  $\hat{G}_{\rm lin}t_{\rm f}$ for any of the two approaches, such that only   elements  above  this level have a conclusive estimation in presence of the current level of shot noise.
  }\label{fig:MS_gate_linear_ ML-LQT}
\end{figure}

As in the single-qubit case, we can first simulate numerically the ideal MS gate subject to shot noise by sampling  $N_{\rm shots}$ times from the ideal probability distribution $
p^{\rm u}_{s,i,\mu}$~\eqref{eq:p_U}. Since the number of shots per configuration $N_{\rm sc}=10^3$ is smaller than in the single-qubit case $N_{\rm sc}=10^4$, one can expect the effect of shot noise to be larger. In fact, we find that the shot-noise threshold for the decay rates is set at $\overline{\gamma}_{\rm sn}=1.7\cdot 10^{-2}$, which is larger than before. As a direct consequence of shot noise, we find that the goodness of fit for the ML-LQT algorithm is worse in this case, reaching chi-square values that still do not reach the convergence criterion $\chi^2_{\rm P,min}=2.1>1=\chi^2_{\rm P,c}$. We depict the estimated Lindblad matrices $\hat{G}_{\rm lin}^{\rm DIA}, \hat{G}_{\rm lin}^{\rm CS}$ for both the   DIA for linear ML-LQT and the CS-LQT in Fig.~\ref{fig:MS_gate_linear_ ML-LQT}. The skyline plot for the linear DIA   displays a fluctuating  and non-sparse landscape (upper panel), with most of the peaks below the percent level being associated to the structureless shot noise. It should be noted that the biggest decay rate that can be extracted in this way is $\hat{\gamma}_{\rm max}=6.7\cdot10^{-2}$, which is less than four times larger than the shot noise threshold, and thus more sensitive to this noise than the trapped-ion single-qubit case, which had a larger number of shots per configuration. The lower panel of Fig.~\ref{fig:MS_gate_linear_ ML-LQT} displays the CS-LQT estimate for the Lindblad matrix, which is also affected by the same level of shot noise, but has a sparser structure so that most of the small peaks associated to shot noise no longer appear.

In addition to the undesired effect of shot noise, one should also mention that there might be errors in the MS gate that go beyond  the LQT assumptions, including SPAM errors of the unitaries used for state preparation and readout, as well as time-dependent and non-Markovian effects that can be expected when considering the dynamical phonons of the two-ion crystal as a small dynamic and thermally-fluctuating environment. In order to discern among these possibilities, insight can be gained by moving to the CS techniques for LQT.
After determining  $\hat{G}$ using the   ML-LQT method, we can diagonalize it and derive a new
basis set $\mathcal{A}$
in which the sparsity of the transformed $G_{\rm true}$ matrix would increase,
which raises the possibility that  CS techniques may actually provide an advantage for LQT. Assuming that the experimental noise only changes mildly in the time between the  ML-LQT an the CS-LQT,   
we can repeat the Lindblad learning  using CS in the sparse basis set $\mathcal{A}$, which would point to a route to make the most of the available number of shots by distributing them among the configurations that carry more information about the relevant noise sources. Since the experimental data~\cite{Pogorelov2021} is already fixed, we will  analyse the prospect of this idea using the same data set, but arranging the configurations so that the ones with more information come first to optimally benefit from CS.

\begin{figure}[!t]
  \centering
  \includegraphics[width=0.9\linewidth]{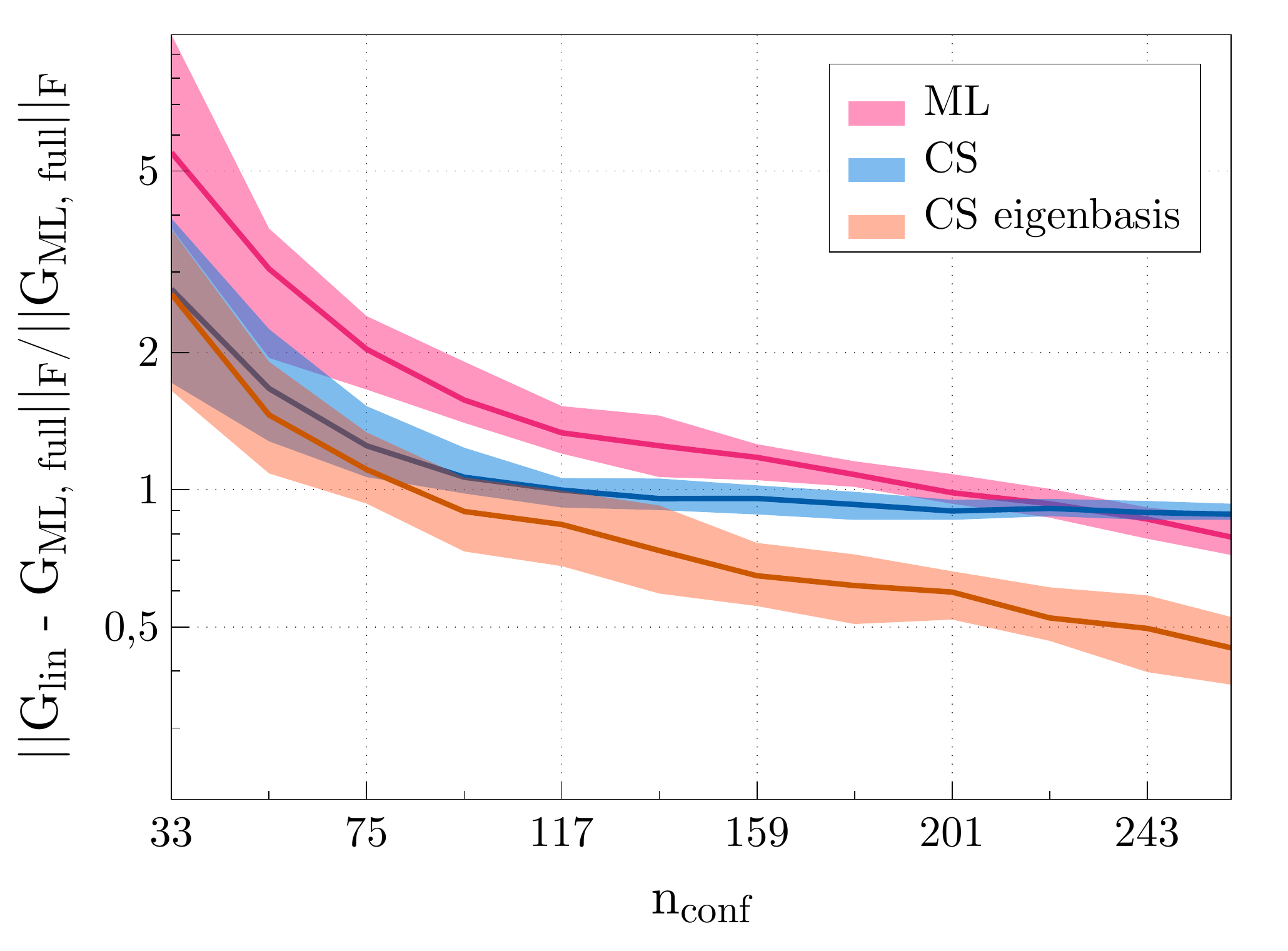}
  \caption{ {\bf  LQT performance for  trapped-ion two-qubit gates:} We apply the full ML-LQT~\eqref{LT_MLE} for $n_{\rm config,i}=432$ and a total of $N_{\rm shots }=1.44\cdot 10^5$ measurements. The estimated  $\hat{G}_{\rm full}$ is used to calculate the Frobenius distances  with respect to the linear estimates based on DIA or and CS learning $\hat{G}_{\rm lin}^{\rm DIA}$ (pink), $\hat{G}_{\rm lin}^{\rm CS}$ (blue), restricted to a smaller number of configurations $n_{\rm config, i}<432$. The specific configurations are selected randomly, and grown in batches of $\delta n_{{\rm config}, i}= 21$ following the approach of Fig.~\ref{fig:LCS_2q_testbed}. Below information completeness at $n_{{\rm config},i}=240$, one can see how the CS-LQT approach presents a clear benefit with respect to the linear ML-LQT. The solid lines represent the median, and the shaded areas the 20/80 percentiles. The orange solid line (shaded area) represents the mean (percentiles) of the CS-LQT using the $\mathcal{A}$ eigenbasis. }
  \label{fig:MS_gate_linear_ ML-LQT_vs_CS}
\end{figure}

In Fig.~\ref{fig:MS_gate_linear_ ML-LQT_vs_CS} we show the results of this analysis.
We start by running the full ML-LQT $\hat{G}_{\textrm{full}}$ considering the informationally complete set of configurations with $n_{\rm config,i}=432$. With this, we can plot the respective normalised Frobenius distances with respect to  the $\hat{G}_{\rm lin}^{\rm DIA}$ or $\hat{G}_{\textrm{lin}}^{\rm CS}$ estimates, as a function of the number of configurations considered, bearing in  mind that there will be an intrinsic error associated to the underlying linear approximation. 
The pink line represents the accuracy of the linear DIA approx, which lies above the CS-LQT  estimate (blue) for small numbers of configurations. We find that only after informational completeness is attained $n_{\rm config,i}=240$, does the linear DIA approach become preferable with respect to the CS-LQT. Finally, we use the information of $\hat{G}_{\rm full}$ to learn about the sparsifying basis $\mathcal{A}$, and adapt the CS-LQT approach by incorporating this information in the linearization~\eqref{eq:lin_contrib_noise}
via a new set of matrices  $\{\mathbb{B}^{pq}\}$~\eqref{B_tensor}. In this way, the Lindblad matrix has a higher sparseness, and  the CS sensing can actually capture more accurately the  relevant noise sources even for a small number of configurations well-below informational completeness. The results displayed in orange  in Fig.~\ref{fig:MS_gate_linear_ ML-LQT_vs_CS} clearly show that the advantage of this idea for the CS-LQT since,  when working in the $\mathcal{A}$ basis, we find that the estimation $\hat{G}_{\rm lin}^{{\rm CS}}|_{\mathcal{A}}$ performs, on average,  considerably better than the previous approaches, always for configuration numbers that  are lower than the informationally-completeness threshold $n_{\textrm{conf}} = 240$.

\section{\bf Conclusions and Outlook}
\label{sec:conclusions}

In this work, we  have introduced and demonstrated  two  improvements for LQT that advance the characterization of current high-fidelity QIPs. First of all, we have shown how one can linearize the ML estimation of the  Lindbladian in the limit of weak noise, which turns the previous non-convex minimizations of LQT~\cite{PhysRevApplied.18.064056,PhysRevApplied.17.054018,PhysRevA.101.062305}  into a convex problem. We have presented a pair of  descent algorithms for linearized ML-LQT that build on the analytical expressions of the gradients of the linear ML estimator, and we have shown that both of these algorithms simplify considerably the classical computational cost of previous full non-convex ML-LQT. We have presented a numerical analysis to show in which situations each of the proposed linear schemes should be preferable, which depends on the rank of the  Lindblad matrix  one wants to  estimate. 

This last observation  led us to develop a compressed sensing approach for LQT, exploiting the structure of the noise, in particular   the existence of leading error sources, to estimate the Lindbladian with a smaller number of configurations, such that one could make the most of the available number of measurement shots. We have shown that CS-LQT can yield important advantages in small QIPs that are limited by shot noise, which could also be extended to larger systems in order to partially minimize the exponential scaling in required resources, provided the   eigenbasis for CS can be extrapolated from the one learnt for the smaller system sizes. We have applied our improved LQT toolbox to experimental trapped-ion data for single- and two-qubit gates considering, for the first time, real non-injected noise. We have shown that our LQT techniques allow us to extract the leading sources of noise, identifying the main Lindblad jump operators responsible for the incoherent noise. For two-qubit entangling gates, we have shown that CS combined with a method to obtain knowledge about the sparsifying basis can be the key to obtain accurate LQT estimates using a fairly small set of measurement shots. 

As an outlook, we believe that it would be interesting to develop  LQT further in two directions. On the one hand, it would be desirable to develop rigorous estimates about the sampling complexity, both for the ML-LQT and CS-LQT strategies discussed in this work, putting them at the same level of other QT protocols for state and process estimation in which the resource scalings are clear, and optimal solutions have been devised. Another interesting avenue for further research is the generalization  of LQT to account for non-Markovian dynamics in open quantum systems. Here, the possibility to include various intermediate times in the evolution $t_i\in T$ will play a crucial role to capture time correlations of the noise. Searching for optimal schemes depending on the structure of the colored noise is an interesting open question.


\acknowledgments
\label{sec:acknowledgments}

We would like to thank the Innsbruck trapped-ion group  for providing us with the experimental data~\cite{Schindler2013, Schindler2013b,Pogorelov2021} used in Sec.~\ref{sec:06_ions}. The project leading to this publication has received funding from the US Army Research Office through Grant No. W911NF-21-1-0007. 
M.M. and L.C. acknowledge support by the ERC Starting Grant QNets Grant Number 804247, the EU H2020-FETFLAG-2018-03 under Grant Agreement number 820495, by the Germany ministry of science and education (BMBF) via the VDI within the project IQuAn, by the Deutsche Forschungsgemeinschaft through Grant No. 449905436, and and under Germany’s Excellence Strategy ‘Cluster of Excellence Matter and Light for Quantum Computing (ML4Q) EXC 2004/1’ 390534769. A.B.~acknowledges support from PID2021-127726NB- I00 (MCIU/AEI/FEDER, UE), from the Grant IFT Centro de Excelencia Severo Ochoa CEX2020-001007-S, funded by MCIN/AEI/10.13039/501100011033, from the CSIC Research Platform on Quantum Technologies PTI-001. A.B. and M.M. acknowledge support  from the European Union’s Horizon Europe research and innovation programme under grant  No 101114305 (“MILLENION-SGA1” EU Project).
The authors gratefully acknowledge the computing time provided to them at the NHR Center NHR4CES at RWTH Aachen University (project number p0020074). 
This is funded by the Federal Ministry of Education and Research, and the state governments participating on the basis of the resolutions of the GWK for national high performance computing at universities (www.nhr-verein.de/unsere-partner).

\appendix

 \section{\bf Complexity of quantum tomography}
\label{app:QT_complexity}

 This appendix serves to set our notation following that of~\cite{velazquez2024dynamical}, and to review key QT  results for the characterization of QIPs. We will devote special attention to understanding the complexity of the various 
 QT strategies, which will serve to frame more precisely our work on Lindbladian tomography.

 In the context of QT, one typically starts from the characterization of quantum states in  systems with a single register of $N$ qubits, which can be repeatedly prepared in a  specific  state $\rho$. This state, which we aim at estimating from measurement data,   is mathematically described by  a positive semidefinite operator of   unit trace  $\rho\in \mathsf{D}(\mathcal{H}_{\rm S})$~\cite{watrous_2018}, where  $\mathcal{H}_{\rm S}=\mathbb{C}^{2}\otimes\cdots^{\hspace{-1.6ex}N}\,\otimes\mathbb{C}^2$ is the $d=2^N$ dimensional $N$-qubit   Hilbert space. By  repeated preparation, we have  $n$ copies of the state  at our disposal, which quantify our resources by fixing the number of samples $N_{\rm shots}=n$ of an underlying probability distribution from which we aim at inferring $\rho$. These copies are thus  measured sequentially for QT according to  a positive operator-valued measure (POVM)~\cite{Chuang1997}, which is  defined by a collection of $|\mathbb{M}_f|$ POVM elements  associated to the measurement outcomes $\{M_\mu:\mu\in\mathbb{M}_f\}$. Each  of these POVM elements   corresponds to a  positive semidefinite operator $M_\mu\in\mathsf{Pos}(\mathcal{H}_{\rm S})$ that acts on an individual copy of the state $\rho$, and the set is constrained to  resolve the identity $\sum_\mu M_\mu=\mathbb{1}_d$~\cite{watrous_2018}. Mathematically, the POVM can be understood as a mapping from  state space to a probability space in which the probability vector $\boldsymbol{p}$ encodes  the full  measurement statistics. According to Born's rule, this mapping reads  \beq
 \label{eq:QST_mapping}
\rho\mapsto\boldsymbol{p}=\sum_\mu{\rm Tr}\{M_\mu\rho\}{\bf e}_\mu\in\mathbb{R}^{|\mathbb{M}_f|},
\eeq
where $\{{\bf e}_\mu:\mu\in\mathbb{M}_f\}$ are the standard unit vectors. The components of $\boldsymbol{p}$ fulfill $p_\mu\geq 0$ and  $\sum_\mu p_{\mu}=1$, which makes the formal connection to  a probability distribution. A measurement is said to be  informationally    complete (IC) if it allows  one to invert the above mapping, and use the measured  probabilities $\boldsymbol{p}$ to recover   the $d^2-1$ real parameters that characterise  a generic quantum state $\rho\in \mathsf{D}(\mathcal{H}_{\rm S})$~\cite{watrous_2018}.  

An IC-POVM  must thus contain    $|\mathbb{M}_f|_{\rm ind}\geq d^{2}$ linearly-independent  elements, exceeding by one the required number of parameters  to estimate $\rho$ due to the constraint $\sum_{\mu}M_\mu=\mathbb{1}_d$~\cite{Flammia2005}.  A standard IC-POVM follows from the  Pauli basis, where we recall that the operators $\big\{E_\alpha:\alpha\in\{0,\cdots, d^2-1\} \big\}$ form an orthogonal basis of the space of linear operators $\mathsf{L}(\mathcal{H}_{\rm S})$  
if ${\rm Tr}\{E_\alpha^\dagger E_\beta\}=d\delta_{\alpha,\beta}\!$, where we consider   the Hilbert-Schmidt scalar product. By taking  tensor products of Pauli matrices, we can define the unormalized Pauli basis   as
\beq
\label{eq:pauli_basis}
E_\alpha\in\mathcal{B}_{\rm P}=\big\{\mathbb{1}_2,\sigma_x,\sigma_y,\sigma_z\big\}^{\!\!\bigotimes^N},
\eeq
 leading to Hermitian and involutory basis operators $E_\alpha^\dagger=E_\alpha, E_\alpha^2=\mathbb{1}_d$, which thus have $m_{\alpha}=\pm 1$ eigenvalues. We can  define a POVM with elements proportional to the orthogonal projectors onto the  corresponding eigenspaces 
\beq
\label{eq:Pauli_basis_proj}
\mu=(\alpha,m_{\alpha}),\hspace{2ex} M_{\alpha,m_{\alpha}}=\frac{1}{d^2-1}\left(\frac{\mathbb{1}_d+m_{\alpha} E_\alpha}{2}\right),
\eeq
where we  excluded $E_0=\mathbb{1}_2{\otimes\cdots\otimes}\mathbb{1}_2=\mathbb{1}_d$ for $\alpha=0$.  The  IC-POVM thus contains $|\mathbb{M}_f|=|\mathbb{M}_\alpha\times\mathbb{M}_{\sigma}|=2(d^2-1)$  elements, and one typically speaks of $d^2-1$ measurement settings with $2$ possible outcomes each. We note that    only $|\mathbb{M}_{\rm ind}|=d^2$   POVM elements  are linearly independent due to the constraint $(d^2-1)\sum_{m_{\alpha}}M_{\alpha,m_{\alpha}}=\mathbb{1}_d$, $\forall\alpha$. 

This measurement setup would allow one to  infer the  non-trivial expectation values  $\langle E_\alpha\rangle=2({d^2-1}){\rm Tr}\{M_{\alpha,+} \rho\}-1 $ that are required for the estimation of the state $\rho$~\cite{nielsen00}. However, except for the single-qubit case, most of these POVM elements  correspond to joint operators  not  native in most QIPs, e.g.  $M_{\mu}\propto(\mathbb{1}_4+\sigma^x\otimes\sigma^x)$ is a projector onto the even-parity  subspace of  $N=2$ qubits,
 which typically requires entangling CNOT gates    preceding an ancilla-qubit projective measurement~\cite{nielsen00}.  To avoid introducing ancillas,  which would also need their own characterization, and  the additional  complexity of working with sequences of entangling gates,  we focus  on a different IC-POVM    with  local Pauli projectors
\beq
\label{eq:local_pauli_proj}
\mu=(\boldsymbol{b},\boldsymbol{m}_{\boldsymbol{b}}),\hspace{2ex}M_{\boldsymbol{b},\boldsymbol{m}_{\boldsymbol{b}}}=\frac{1}{3^N}P_{b_1,m_{b_1}}\otimes\cdots\otimes P_{b_N,m_{b_N}}.
\eeq
Here,  $b_j\in\{x,y,z\}$ indicates the Pauli basis of the  $j$-th qubit projector,  and $m_{b_j}=\pm 1$ the corresponding eigenvalue, e.g. $P_{x,\pm}=(\mathbb{1}_2\pm\sigma_x)/2$. Altogether,  we have $3^N$ measurement bases with $d$ possible outcomes each. Therefore, the POVM has  $|\mathbb{M}_f|=|\mathbb{M}_{\boldsymbol{b}}\times\mathbb{M}_{\boldsymbol{m}_{\boldsymbol{b}}}|=3^N d$ elements, although  not all of them  are independent, as  the set of projectors with a fixed  basis  resolves the identity $\sum_{\boldsymbol{m}_{\boldsymbol{b}}}P_{(b_1,m_{b_1})}\otimes\cdots \otimes P_{(b_N,m_{b_N})}=\mathbb{1}_d$.  Accordingly, we  have $|\mathbb{M}_{\rm ind}|=3^N(d-1)+1\geq d^2$ independent  POVM elements and, thus,  an IC-POVM. 

According to the QT scenario described above, one would  ideally need to prepare the same  state $\rho$ repeatedly and measure it  a number of times that scales with either $4^N$ times for global Pauli measurements~\eqref{eq:Pauli_basis_proj} or  $6^N$ times for local ones~\eqref{eq:local_pauli_proj}.
In practice, however, any of these resource counts is an idealization, as  various  sources of noise and errors  occur in any  experiment, rendering the  measurements imperfect. At the very least, one is always confronted with projection/shot noise due to the finite number of measurement shots $N_{\boldsymbol{b}}$ per measurement basis~\cite{PhysRevA.47.3554}, which can only provide us with a relative-frequency approximation of the probability vector 
\beq
\label{eq:state_QT_finite_fre}
\boldsymbol{f}=\sum_{\boldsymbol{b},\boldsymbol{m}_{\boldsymbol{b}}} \frac{N_{\boldsymbol{b},\boldsymbol{m}_{\boldsymbol{b}}}}{N_{\rm shots}}\,{\bf e}_{\boldsymbol{b},\boldsymbol{m}_{\boldsymbol{b}}}\approx\boldsymbol{p}, \hspace{2ex}\sum_{\boldsymbol{b},m_{\boldsymbol{b}}} f_{{\boldsymbol{b},\boldsymbol{m}_{\boldsymbol{b}}}}=1.
\eeq
Here, $N_{\boldsymbol{b},\boldsymbol{m}_{\boldsymbol{b}}}$ stands for the number of observed $\boldsymbol{m}_{\boldsymbol{b}}$  outcomes associated to the measurement basis $\boldsymbol{b}$, such that $N_{\boldsymbol{b}}=\sum_{\boldsymbol{m}_{\boldsymbol{b}}}N_{\boldsymbol{b},\boldsymbol{m}_{\boldsymbol{b}}}$ and $N_{\rm shots}=\sum_{\boldsymbol{b}}N_{\boldsymbol{b}}$ is the total number of shots in the experiment and, thus, the number of copies of the state $\rho$ that the register must be sequentially prepared into $N_{\rm shots}=n$.
Hence,  a more realistic description should account for errors in the  inversion of Eq.~\eqref{eq:QST_mapping}, which can lead to the estimation of unphysical states when dealing with overcomplete POVMs,  require one to move to  maximum-likelihood methods. 

In general,   QT can  only aim at estimating the state $\rho$ by an approximate $\hat{\rho}$  with a certain  error $\epsilon>0$. This error can be quantified by the trace-distance of the corresponding states  $\epsilon=\parallel\!\!\rho-\hat{\rho}\!\!\parallel_{1}={\rm Tr}\big\{\sqrt{(\rho-\hat{\rho})^2}\big\}$, where one uses   the Schatten 1-norm~\cite{watrous_2018}. Therefore, a more meaningful question regarding the  complexity of state-QT is  to quantify the  resources required to reach a desired target error $\epsilon$. In the present context, this should be expressed in terms of the required number of copies  of the state at our disposal, which also correspond to the total number of measurement shots that will be performed $N_{\rm shots}$, and will be larger than the previous idealized scalings. As shown in~\cite{KUENG201788,7956181,chen2022tight,comment}, the optimal strategy using a fixed sequence of POVM elements employs  randomised measurements,  which are obtained by acting with $N_{\rm shots}$ uniformly-sampled random  unitaries $U\in\mathsf{U}(\mathcal{H}_{\rm S})$ prior to a sequence of specific projective measurements of the resulting states on the computational basis. By finding  both upper and lower  bounds on the resources, these works show that this  scheme  is optimal and requires  $N_{\rm shots}\propto 2^{3N}/\epsilon^2$ copies of the state~\cite{KUENG201788,7956181,chen2022tight}. Note that a better scaling $N_{\rm shots}\propto2^{2N}/\epsilon^2$ can be achieved if one has access to a multiple-copy register $\rho^{\otimes N_{\rm shots}}$ and can perform collective entangled measurements~\cite{7956181,10.1145/2897518.2897544}. 
Finally, the sampling complexity  can change if the number of outcomes of each of the projective measurements is independent of the system size. An important example is that of the  binary-outcome  Pauli measurements in Eq.~\eqref{eq:Pauli_basis_proj}, where the scaling is instead  $N_{\rm shots}\propto 2^{4N}/\epsilon^2$ and is also  optimal~\cite{Lowe_Angus2021}. 

Due to this  exponential sampling complexity, the resources for the standard approach to state-QT become prohibitive already for intermediate-sized systems. As discussed in the introduction, there has been a considerable effort in devising alternative  schemes with a lower cost by e.g. restricting the set of possible states to an ansatz with a smaller number of parameters. This  can be motivated by  a specific  symmetry, such as permutation symmetry~\cite{dariano_23,PhysRevLett.105.250403,Moroder_2012}, which leads to a polynomial scaling with $N_{\rm shots}$. Instead, one can restrict the set of possible states  according to their entanglement content, and perform efficient QT within the set of area-law states employing polynomial resources~\cite{Cramer2010,PhysRevLett.111.020401,Lanyon2017}. A different assumption is that of low-rank QT which, although  allowing for a smaller gain,  is less restrictive and, arguably, common to the majority of states created in recent experiments. With the low  errors achieved by current QIPs, these states  are close to  ideal pure states $\rho\approx\ket{\psi}\!\!\bra{\psi}$ and, thus,    have  a low rank $r\ll 2^{N}$. Building on ideas of compressed sensing to recover a large sparse vector or matrix by randomly sampling a much smaller number of its elements~\cite{Donoho2006, Candes2006},  compressed sensing QT has been shown to require a number of copies $N_{\rm shots}$ that scales with $\mathcal{O}( Nr^2 2^{2N}/\epsilon^2)$ when based on Pauli measurements~\cite{PhysRevLett.105.150401,Flammia_2012}. This scaling can be improved further to $\mathcal{O}(r^2 2^{N}/\epsilon^2)$ by considering collective measurements on the multiple-copy register $\rho^{\otimes N_{\rm shots}}$~\cite{KUENG201788,7956181,7956181}.  Although these sampling complexities still scale exponentially with the number of qubits $N_{\rm shots}$, the improvement is considerable in comparison to the previous  scaling $N_{\rm shots}\propto 2^{4N}/\epsilon^2$ in  standard  state-QT.

Let us now discuss the complexity of QT for the time evolution of a quantum system, commonly referred to  as quantum process tomography~\cite{doi:10.1080/09500349708231894,PhysRevLett.78.390,PhysRevA.63.020101,PhysRevA.63.054104,PhysRevA.68.012305}. This  task  is also of primary importance in QIPs for which the precise characterization of a  universal gate set~\cite{Chuang1997} allows one to identify, model, and possibly amend the errors in a quantum computation. Knowing the precise error model of the native gate set is important for an accurate estimation of the error threshold in fault-tolerant quantum computation~\cite{10.1063/1.1499754,Raussendorf_2007,PhysRevA.80.052312,landahl2011faulttolerant,PhysRevLett.103.090501}. In an idealised error-free situation, the  exponential scaling of resources follows directly from the description of any admissible quantum evolution  by a dynamical quantum map $\mathcal{E}_{t,t_0}\in\mathsf{C}(\mathcal{H}_{\rm S})$ $\forall t\in T=[t_0,t_{\rm f}]$.  Each snapshot of the dynamical quantum map belongs to the set of  completely-positive and trace-preserving channels $\mathsf{C}(\mathcal{H}_{\rm S})$ acting on  the  space of linear operators $\mathsf{L}(\mathcal{H}_{\rm S})$~\cite{watrous_2018}. In the following discussion, we will only refer to  the tomography of a quantum channel, and hence consider a single snapshot at $t\in T$. This channel admits a representation in terms of a  process $\chi$ matrix
\beq
\label{eq:chi_matrix}
\rho_0\mapsto\rho(t)=\mathcal{E}_{t,t_0}(\rho_0)=\sum_{\alpha,\beta}\chi_{\alpha\beta}(t,t_0)E_\alpha^{\phantom{\dagger}}\rho_0 E_{\beta}^{\dagger}.
\eeq
 For Eq.~\eqref{eq:chi_matrix} to represent an admissible physical process,  the   process matrix  $\chi(t,t_0)$ must be semidefinite positive   and subjected to a so-called~\cite{Chuang1997} trace constraint 
 \beq
 \label{eq:trace_constraint}
 \chi_{\alpha\beta}(t,t_0)\in\mathsf{Pos}(\mathbb{C}^{d^2}),\hspace{1ex} \sum_{\alpha,\beta}\chi_{\alpha\beta}(t,t_0)E^\dagger_\beta E_\alpha^{\phantom{\dagger}}=\mathbb{1}_d.
 \eeq
  Accordingly,  $d^2(d^2-1)$ real parameters per snapshot are required  to describe the time evolution  $\rho(t)\in\mathsf{D}(\mathcal{H}_{\rm S})$,  and one will similarly to state-QT face exponential scalings~\cite{PhysRevA.77.032322}. 
  
    In order to determine the time evolution of  $\chi(t,t_0)$,  the standard process-QT~\cite{doi:10.1080/09500349708231894,PhysRevLett.78.390} requires the preparation of $|\mathbb{S}_0|=d^2$ linearly-independent initial states $\{\rho_{0,s}:s\in\mathbb{S}_0\}$. A typical choice is to consider all possible tensor products of four  states
    \beq
    \label{eq:IC_initial_set}
    \rho_{0,s}\in\big\{\ket{0}\!\!\bra{0},\ket{1}\!\!\bra{1},\ket{+}\!\!\bra{+},\ket{+{\rm i}}\!\!\bra{+{\rm i}}\big\}^{\!\otimes^N},
    \eeq    
    where $\{\ket{0},\ket{1}\}$ are the qubit computational basis, and $\{\ket{+}=(\ket{0}+\ket{1})/\sqrt{2},\ket{+\ii}=(\ket{0}+\ii\ket{1})/\sqrt{2}\}$ are two other cardinal states on the qubit Bloch sphere.
    Additionally,  one must probe the system with an IC-POVM measurement  at various instants of time, each of which is  described by  $|\mathbb{M}_{\rm ind}|\geq d^{2}$ linearly-independent  elements $\{M_\mu:\mu\in\mathbb{M}_f\}$, e.g. the  IC-POVM in Eq.~\eqref{eq:local_pauli_proj}. One thus gets a formal mapping between the process matrix and a probability matrix 
    \beq 
    \chi(t,t_0) \mapsto \big[p(t,t_0)\big]_{\mu s}={\rm Tr}\big\{M_\mu\mathcal{E}_{t,t_0}(\rho_{0,s})\big\},
    \eeq
    the columns of which correspond to the probability vectors for the measurement statistics. To recover the full dynamical quantum map, this equation must be inverted~\cite{doi:10.1080/09500349708231894,PhysRevLett.78.390} at each instant of time $\{t_i:i\in\mathbb{I}_t\}\subset T$, which has an additional overhead increasing  the QT complexity  even further. The independent triples $(\rho_{0,s},t_i,M_{\mu})$ form our LQT configurations. Considering the set of initial states~\eqref{eq:IC_initial_set}, and the  independent   POVM projectors used in our work~\eqref{eq:local_pauli_proj}, we thus have 
    \beq 
    \label{eq:n_conf}
    n_{{\rm conf},i}=3^Nd^2(d-1)
    \eeq
    configurations per time step.  Each of this requires preparing a single copy of the time-evolved states, such that the resources in terms of the number of configurations is  $n_{\rm config}=\sum_i n_{{\rm conf,}i}$. In QPT, one typically focuses on reconstructing the quantum channel for a  single snapshot $t_i\in T$. 

In a realistic scenario, one must again consider the estimation  error $\epsilon$, which will require repeating the measurements a total number of times $N_{\rm shots}$ that is much larger than the above configurations. Due to the finite number of repetitions, shot noise only provides us with an approximation to the above probability matrix which, using Eq.~\eqref{eq:local_pauli_proj}, reads
\beq
\label{eq:process_QT_finite_fre}
f(t_i,t_0)=\sum_{s}\sum_{\boldsymbol{b},\boldsymbol{m}_{\boldsymbol{b}}} \frac{N_{s,i,\boldsymbol{b},\boldsymbol{m}_{\boldsymbol{b}}}}{N_{s,i,  \boldsymbol{b}}}\,{\bf e}_{\boldsymbol{b},\boldsymbol{m}_{\boldsymbol{b}}}\otimes {\bf e}_s\approx{p}(t,t_0). 
\eeq
Here, $N_{s,i,\boldsymbol{b},\boldsymbol{m}_{\boldsymbol{b}}}$ stands for the number of observed $\boldsymbol{m}_{\boldsymbol{b}}$-outcomes associated to each  measurement basis $\boldsymbol{b}$, and now also to each initial state $\rho_{0,s}$ for the specific snapshot $t_i\in T$. Hence, $N_{s,i,\boldsymbol{b}}=\sum_{\boldsymbol{m}_{\boldsymbol{b}}}N_{s,i, \boldsymbol{b},\boldsymbol{m}_{\boldsymbol{b}}}$ is the number of shots per initialization  and measurement basis, and $N_{\rm shots}=\sum_{s,i,\boldsymbol{b}}N_{s,i, \boldsymbol{b}}$ is the total number of shots performed for the process-QT.  

In general,  the rigorous proofs that underlie our previous discussion on the  sampling complexity of state-QT    cannot be directly ported   to  process-QT~\cite{Kliesch2019guaranteedrecovery}, except  in some particular cases. Paralleling our description of state-QT, one can restrict the  processes $\mathcal{E}_{t,t_0}$ to specific types, in search for more efficient  strategies for process-QT. For instance, if the process is restricted to  the family of Pauli channels, one can devise almost optimal QT strategies, and even provide mathematical proofs addressing the sample complexity~\cite{10.1145/3408039,Harper2020}. In this type of channels, one considers the above Pauli basis $E_\alpha\in \mathcal{B}_{\rm P}$~\eqref{eq:pauli_basis} as the   orthonormal operator basis in Eq.~\eqref{eq:chi_matrix}, and the channel ansatz is restricted by imposing that the process matrix must be diagonal $\chi_{\alpha\beta}(t,t_0)=p_{P,\alpha}(t-t_0)\delta_{\alpha,\beta}$, where $\delta_{\alpha,\beta}$ is the Kronecker delta. Then, the positive semidefinite constraint simply requires the positivity of ${p}_{P,\alpha}\geq 0$, and  the trace constraint becomes  $\sum_\alpha p_{P,\alpha}=1$,  such that  the channel is fully determined by a probability vector $\boldsymbol{p}_P(t-t_0)$ with components  known as the Pauli error rates. Process-QT then requires estimating the vector of Pauli error rates  by an approximate one $\hat{\boldsymbol{p}}(t-t_0)$ with a given  $1$-norm distance $\epsilon=\parallel\!\!\boldsymbol{p}_P(t-t_0)-\hat{\boldsymbol{p}}_P(t-t_0)\!\!\parallel_1=\sum_\alpha|p_{P,\alpha}(t-t_0)-\hat{p}_{P,\alpha}(t-t_0)|$. 

In a similar spirit  to the previous QT scenario, one may consider a quantum register of $N_{\rm shots}$ qubits that can be initialized in different  states, then  evolved under the channel that we aim at estimating, which may also be interleaved with a fixed sequence of unitary gates. Finally, each of the $N_{\rm shots}$  resulting states can be measured individually with a fixed  sequence of POVM elements. In this case, we do not have  exact copies of the same state as in state-QT, which are probed by distributing the finite number of $N_{\rm shots}$ experimental shots among the smaller number of measurement settings, but rather  a collection of configurations  involving  different time-evolved states  measured in different  settings, among which  we distribute the total number of shots $N_{\rm shots}$. As discussed in~\cite{10.1145/3408039,fawzi2023lower}, when the channel to be estimated is a snapshot $\mathcal{E}_{t,t_0}$  of the Pauli type, and one uses intermediate random Pauli gates to average over the noise, the resources scale with $N_{\rm shots}=\mathcal{O}(N2^{3N}/\epsilon^2)$ which, up to a logarithmic correction, is similar to the optimal scaling for the QT of states~\cite{KUENG201788,7956181,chen2022tight}. Indeed, a lower bound presented in~\cite{fawzi2023lower} shows that, up to these logarithmic corrections, this  Pauli-channel tomography is already  optimal. It is interesting to note that performing Pauli tomography can actually lead to quantum advantage, as  the  complexity  using a quantum processor can lead to an exponential reduction~\cite{PhysRevA.105.032435}. 

Similarly to our discussion of quantum state tomography, we now discuss a less restrictive assumption  that  allows  applying compressed sensing for the  QT of channels~\cite{kosut2009quantum,PhysRevLett.106.100401,Flammia_2012, PhysRevA.90.012110,Kliesch2019guaranteedrecovery}. The process-matrix description~\eqref{eq:chi_matrix} is equivalent to the Kraus operator-sum representation~\cite{Hellwig1970,KRAUS1971311}, namely
\beq
\label{eq:Krauss}
\mathcal{E}_{t,t_0}(\rho_0)=\sum_{n}K_n^{\phantom{\dagger}}(t,t_0)\rho_0 K_{n}^{\dagger}(t,t_0),
\eeq
where the Kraus operators are constrained to  resolve the identity in order to describe a completely-positive trace-preserving map.  Indeed, this constraint can be obtained by constructing the Kraus operators as follows 
\beq 
\label{eq:Kraus_constraint}
\sum_nK_{n}^{\dagger}(t,t_0)K_{n}^{\phantom{\dagger}}(t,t_0)=\mathbb{1}_d, \hspace{1ex} K_n^{\phantom{\dagger}}(t,t_0) = \sqrt{\chi_n}\sum_{\alpha} {{v}}_{\alpha, n}E_\alpha,
\eeq
where $\chi_n$ and $\boldsymbol{v}_n=\sum_\alpha{{v}}_{\alpha, n}{\bf e}_\alpha$ are the eigenvalues and eigenvectors of the process matrix, $\chi(t,t_0)\boldsymbol{v}_n=\chi_n\boldsymbol{v}_n$, respectively. For some quantum channels, it may turn out that the process matrix has zero eigenvalues, such that the sum in the Kraus decomposition~\eqref{eq:Krauss} is truncated
\beq
\label{eq:Krauss_trunc}
\mathcal{E}_{t,t_0}(\rho_0)\approx\sum_{n=1}^{r_\kappa}K_n^{\phantom{\dagger}}(t,t_0)\rho_0 K_{n}^{\dagger}(t,t_0),
\eeq
terminating at a certain Kraus rank $r_{\kappa}\leq d^2=4^N$. This is the case, for instance, of   pure-unitary dynamics in which   $r_{\kappa}=1$. 

Just as the QT of most states produced by current QIPs can be improved by assuming low-rank states, the gates that these processors employ to produce such states are very close to unitaries, and thus have a low Kraus rank $r_{\kappa}\ll 4^N$. As discussed in~\cite{Flammia_2012,Kliesch2019guaranteedrecovery}, compressed sensing  of low-rank channels using a fixed sequence of state preparation, evolution, and Pauli measurements,  allow one to derive upper bounds on the required resources  $N_{\rm shots}\leq\mathcal{O}(N2^{5N}/\epsilon^2)$. Although, to our knowledge, there are no lower bounds  to  precisely determine the sampling complexity and argue about optimality, the exponential scaling in resources is clear. At this point, it is important to  note that, if one could further restrict the low-rank channel by knowing which specific unitary operator the time evolution is  close to, which is typically the case of  high-fidelity QIPs, one could use a specific basis where the process matrix is sparse, and obtain further improvements for QT. In fact, if one has this prior information, although there are still no strict upper and lower bounds for  the sample complexity, the number of QT configurations has been shown to scale polynomially~\cite{PhysRevLett.106.100401,Rodionov2014}.

When the dynamics is purely unitary, instead of estimating the quantum channels for various snapshots to reconstruct a coarse-grained version of the dynamical quantum map, one may directly target the system Hamiltonian $H$ generating such a time evolution. Indeed, for dynamical quantum maps~\eqref{eq:Krauss} with Kraus rank $r_\kappa=1$,  the constraint~\eqref{eq:Kraus_constraint} can be fulfilled by considering a single  time-ordered exponential  with a time-dependent Hamiltonian $K_1\!(t,t_0)=\mathcal{T}\big\{{\rm exp}\big(\!-\ii\!\int_{t_0}^{t}\!\!{\rm d}t' H(t')\big)\big\}$, the estimation of which can be less resource intensive, specially when the Hamiltonian is constant~\eqref{eq:spanned_H}. The idea is that the typical Hamiltonians describing physical QIPs do not actually require  $d^2-1$ parameters when one chooses an appropriate operator basis, such as $E_\alpha\in \mathcal{B}_{\rm P}$~\eqref{eq:pauli_basis}. Taking into account the tensor-product character of this basis, and the typical locality of interactions, one can restrict oneself to 
\beq
\label{eq:trunc_H}
H\approx\sum_{\alpha=1}^m c_\alpha E_\alpha.
\eeq
Here, the Hamilonian is parametrized by a  smaller set of  real coefficients  $\{c_\alpha: \alpha\in\{1,\cdots,m \}\} $ with  $m=\mathcal{O}({\rm poly}(N))\ll 4^N$~\cite{PRXQuantum.2.010102}, which clearly resembles the situation that motivated the low-rank truncation  of the quantum channel~\eqref{eq:Krauss_trunc}. Remarkably, if the initial state commutes with the Hamiltonian,
the estimation of the Hamiltonian  only requires  solving a linear system of equations for the vector of couplings $\boldsymbol{c}$, which lies in the kernel of an observable correlation matrix $C$. This matrix consists of equal-time two-point functions in the operator basis $C_{\alpha\beta}={\rm Tr}\{\ii[E_\alpha,E_\beta]\rho\}$~\cite{PhysRevLett.107.210404,PhysRevX.8.031029,Qi2019determininglocal,PhysRevLett.122.020504,PhysRevLett.124.160502,evans2019scalable}. Allowing for adaptive sequences where the initial state,  control fields, and measurements, can be varied  according to a Bayesian update, one can find an estimate $\hat{\boldsymbol{c}}$ considering a fidelity error $F(\hat{\boldsymbol{c}}, \boldsymbol{c})=|\hat{\boldsymbol{c}}^\dagger \boldsymbol{c}|^2/(\parallel\!\! \hat{\boldsymbol{c}}\!\!\parallel_2^2\parallel\!\! \boldsymbol{c}\!\!\parallel_2^2)\leq 1-\epsilon$~\cite{evans2019scalable}.
In the work~\cite{evans2019scalable}, the authors  showed that, up to logarithmic corrections,  the  complexity of this Hamiltonian learning is polynomial $N_{\rm shots}=\mathcal{O}(N^{3}k^{3D}/(\epsilon\Delta)^{3/2})$, where  a $k$-local Hamiltonian in $D$ dimensions has at most $m=\mathcal{O}(Nk^{3D})$ terms, and  the correlation matrix is assumed to have  an energy gap $\Delta$.

As emphasised in the introduction, in order to account for  realistic errors in QIPs, one should upgrade closed-system Hamiltonian QT to  open quantum systems that do not  evolve unitarily. The theory of 
open quantum systems aims at describing the dynamics of the system  after tracing over an ever-present environment $\rho={\rm Tr}_{E}\{\ket{\Psi_{\rm SE}}\!\bra{\Psi_{\rm SE}}\}\in\mathsf{D}(\mathcal{H}_{\rm S})$ with $\ket{\Psi_{\rm SE}}\in\mathcal{H}_{\rm SE}$, which can be generally described by an exact integro-differential   equation known as the Nakajima-Zwanzig  equation~\cite{10.1143/PTP.20.948,Zwanzig1960EnsembleMI,Breuer2002}. When the coupling to the environment is weak, and the timescale of interest   is much larger than the environmental correlation time $\tau_{\rm c
}$, one can approximate it by a much simpler Markovian master equation. In more detail,  $\tau_{\rm c}$ sets a characteristic timescale for the decay of the environmental correlations and, when the time-scale of interest is much larger $\Delta t\gg \tau_{\rm c}$, one can  coarse-grain  to derive a time-local differential equation for the quantum system that is not affected by environmental memory effects. In this Born-Markov limit, this equation  has a Lindblad form~\eqref{eq:lindblad_generator}, such that ${{\rm d}\rho}/{{\rm d}t} =\mathcal{L}(\rho)$ is fully determined by the Hamiltonian ${H}$ and Lindblad matrix $G$. The goal of Lindbladian QT is to estimate both the Hamiltonian $H$ and  Lindblad matrix $G$~\eqref{eq:lindblad_generator}. Let us note that, by choosing the Pauli basis with $E_0=\mathbb{I}_{d}$, one can incorporate some of the Lindbladian contributions in the Hamiltonian part $H\mapsto\tilde{H}=H +\sum_\alpha {\rm Im}\{G_{0\alpha}\}E_\alpha $, such that the fully incoherent part of the dynamics is encapsulated in a $({d^{2}-1})\times ({d^{2}-1})$  positive semidefinite matrix $G\to\tilde{G}\in\mathsf{Pos}(\mathbb{C}^{d^2-1})$. In the main text, we avoid the tildes to simplify notation.
Hence, in addition to the $d^{2}-1$ real parameters for the Hamiltonian $\{\tilde{c}_\alpha=c_\alpha+{\rm Im}\{G_{0\alpha}\}\}$, we require an additional set of  $(d^2-1)^2$ real parameters for  $\{G_{\alpha\beta}\}$. Altogether this yields $d^2(d^2-1)$, which is the same parameter count one finds for the full process matrix.

\section{\bf Linearizing Lindbladian quantum tomography}
\label{app:app_linearization}

In this Appendix, we elaborate on the linearization of the Markovian evolution operator $\exp((t-t_0)\mathcal{L} )$, following an approach based on error process matrices~\cite{Korotkov2013}. The main idea in~\cite{Korotkov2013} is to factor out the target unitary  $U(t,t_0)$ in the estimation of the process matrix $\chi(t,t_0)$ of a noisy gate~\eqref{eq:chi_matrix}, and to only learn the error process matrix $\chi^{\rm err}(t,t_0)$. As discussed in the main text, in the limit of weak noise, this  factoring can have important practical consequences in ML-LQT and CS-LQT, rendering the constrained optimization convex and the Lindbladian estimation much more efficient.
We assume  knowledge of the coherent part of the evolution $U(t,t_0)={\rm exp}(-\ii (t-t_0) H)$, which narrows the learning task to the  sole estimation of the  Lindblad matrix $G$.
The state of the system at a  time $t_{i+1}=t_i+\Delta t$ can be derived from the Lindblad master equation,  which, 
in  the weak-noise limit, can be approximated by a Suzuki-Trotter expansion. Moreover,   the dissipative evolution operator now admits a linear approximation $\exp \left(\mathcal{L}_G\Delta t\right) \approx \left(\mathsf{I}+\mathcal{L}_G\Delta t\right)$ even when the coherent part
is not small, where $\mathsf{I}(\rho)=\rho$ is the identity channel.
Preparing the ground for the linearized compressed-sensing formulation, we will allow the Lindbladian to be defined as
\begin{equation}
    \label{eq:lindblad_generator_app}
    \mathcal{L}_G(\rho) =\sum_{p,q}\! G_{pq}\!\! \left( B_p^{\phantom{\dagger}} \rho B_q^{\dagger} - \half \big\{ B_q^{\dagger} B_p^{\phantom{\dagger}}, \rho \big\} \!\right)\!,
\end{equation}
 where we consider any operator basis $\mathsf{L}(\mathcal{H}_{\rm S})={\rm span}\{\mathcal{B}\}$ with  $\mathcal{B}=\{B_p:p\in\{0,\cdots, d^2-1\}\}$, and only impose that ${\rm Tr}\{B_p\}=0$. To simplify the presentation, we keep the same notation of the $G$ matrix as in the Pauli-basis Eq.~\eqref{eq:lindblad_generator}, although the main idea is that the matrix will be different and sparser if we allow the $B_p=\sum_{p}b^{p}_{\alpha}E_\alpha$ to be certain linear combinations of Pauli operators. A clear and simple example is that of a single-qubit   spontaneous emission where using $B_p=(\sigma_x-\ii\sigma_y)/2=\sigma^-$   can increase sparseness. 

The equation for time evolution from the $s$-th initial state can then be expanded as 
\begin{widetext}
\begin{align}
    \rho_s( t_{i+1})
        \approx U(t_{i+1},t_i)\rho_s(t_i) U^\dag\!(t_{i+1},t_i) + U(t_{i+1},t_i) \sum_{p,q}G_{p,q}\Delta t\sum_{\alpha,\beta}b^p_\alpha b^{q*}_\beta  \left(E_\alpha^{\phantom{\dag}}\rho_s(t_i) E_\beta^\dag - \frac{1}{2}\big\{E_\beta^\dag E_\alpha^{\phantom{\dagger}},\rho_s(t_i)\big\}\right) U^\dag\!(t_{i+1},t_i).
\end{align}
\end{widetext}
This leading-order  evolution can be rewritten  compactly as
\beq
\rho_s( t_{i+1}) \approx \sum_{\alpha,\beta}\bigg(\chi_{\alpha\beta}^I + \sum_{\gamma,\delta}\mathbb{B}^{p q}_{\alpha\beta}G_{pq}\Delta t\bigg)U_{\Delta t}E_\alpha^{\phantom{\dag}}\rho_s(t_i) E_\beta^\dag U^\dag_{\Delta t},\,
\eeq
where we have introduced a short-hand notation $U_{\Delta t}=U(t_{i+1},t_i)$, and the following quantities following from an  expansion in the Pauli basis and standard algebra
\begin{align}
\label{B_tensor}
  &\chi_{\alpha\beta}^I = \delta_{\alpha,0}\delta_{\beta,0} \,,\hspace{2ex}
  \mathbb{B}^{p q}_{\alpha\beta} = b^{p}_{\alpha}b^{q}_{\beta}
	- \half\big(c^{p q}_\beta \delta_{\alpha,0} + \delta_{0,\beta}c^{pq*}_\alpha\big), 
\end{align}
which also depend on  
\beq
\label{c_tensor}
 c^{p q}_{\alpha} = \sum_{\gamma,\delta}b^{p}_{\gamma}b^{q*}_{\delta}\text{Tr}
  \big\{E^\dag_\alpha E^\dag_\delta E_\gamma^{\phantom{\dagger}}\big\}.
  \eeq
We can now make use of Eq.~\eqref{eq:chi_matrix} to identify  a  process  matrix for the infinitesimal dynamical quantum map in the weak-noise limit. The only difference is that the state resulting from this map is then acted with the unitary  as follows
\begin{equation}
 \rho_s(t_{i+1})\approx U_{\Delta t}\bigg(\sum_{\alpha,\beta}{\chi}^{\rm err}_{\alpha\beta}(t_{i+1},t_i)E_\alpha\rho_s(t_i)E_{\beta}\bigg)U^{\dagger}_{\Delta t}\,.
  \label{eq:chi_delta_t}
\end{equation}
Since we have factored out the effect of the ideal target  unitary $U(t_{i+1},t_i)$ from deviations caused by the dissipation, this process matrix actually contains information about the noise and is known as the infinitesimal error matrix~\cite{Korotkov2013,PhysRevLett.108.057002}, namely 
\beq
{\chi}^{\rm err}_{\alpha,\beta}(t_{i+1},t_i)=\chi_{\alpha\beta}^I + \sum_{\gamma,\delta}\mathbb{B}^{p q}_{\alpha\beta}G_{pq}\Delta t.
\eeq
 The novelty in our work with respect to~\cite{Korotkov2013} is that, as noted above,  we allow for a more general general basis, and will consider this linearization in the context of ML estimation.

The contribution from the first term $\chi^I$ actually leads to the ideal unitary, whereas the second term $\Delta{\chi}^{\rm err}(t_{i+1},t_i)$ is responsible for the small errors in the weak-noise limit.
The full dynamical quantum map for the time evolution $t\in T$  can be obtained by composing the above   infinitesimal maps for small time periods $\Delta t$ to obtain the following temporal sequence of $M$ channels with $\Delta t=(t_{\rm f}-t_0)/M$ and $t_i=t_0+(\Delta t) i$. This can be schematically depicted as follows
\begin{widetext}
\begin{equation}
\label{eq:chi_sequence}
  \text{---} \; {\chi}^{\rm err}(t_{1},t_0) \;\text{---} \; U(t_{1},t_0) \;\text{---}
         	\; {\chi}^{\rm err}( t_2,t_1)\;\text{---} \; U(t_{2},t_1) \;\text{---}\,\ \dots
  \,\,\text{---} \; {\chi}^{\rm err}(t_{M},t_{M-1}) \;\text{---} \; U(t_{M},t_{M-1}) \;\text{---}\,,
\end{equation}
\end{widetext}
As we can see, the dynamical quantum map is described by the  composition of, first, the infinitesimal error channel, followed by the infinitesimal unitary  evolution, which is then repeated sequentially. In most theoretical treatments of noisy circuits, however, one models a faulty gate by the action of the full ideal unitary composed with the  the full error channel. To make connection with these studies, we can swap the order of any unitary $V$ and $\tilde{\chi}^{\rm err}$ by a simple unitary transformation
\begin{equation*}
  \text{---} \; V \;\text{---} \; {\chi}^{\rm err} \;\text{---}\;\; =
  \;\;
  \text{---} \; {\chi}^{\rm err}_{V} \;\text{---} \; V \;\text{---}\;\;
\end{equation*}
where we have introduced the following basis transformation for the error process matrix  
\begin{equation}
  \chi^{\rm err} = \mathbb{W}^\dag {\chi}^{\rm err}_{U_{{\Delta t}}} \mathbb{W}\,,\;\hspace{2ex}
  \mathbb{W}_{\alpha\beta} = \text{Tr}\left\{E_\alpha^\dag U^{\phantom{\dagger}}_{\Delta t}E_\beta^{\phantom{\dagger}} U^\dag_{\Delta t}\right\}\,.
\end{equation}
With this transformation, we can swap all $\chi^{\rm err}(t_{i+1,t_i})$ in Eq.~\eqref{eq:chi_sequence} to the left, such that the dynamical quantum map is described by
\begin{widetext}
\begin{equation*}
  \text{---} \; \chi^{\rm err}( t_1,t_0) \;\text{---} \; \chi^{\rm err}_{U( t_1,t_0)}( t_2,t_1) \;\text{---}
         	\; \chi^{\rm err}_{U(t_2,t_1)U(t_1, t_0)}(t_3,t_2) \;\text{---} \; \dots \;\text{---}\; U( t_1, t_0)\; \text{---} \; U( t_2, t_1) \;\text{---}U( t_3, t_2) \;\text{---} \; \dots\,.
\end{equation*}
\end{widetext}
In the first order approximation~\cite{Korotkov2013}, the composition of a sequence of error process matrices can be written as a sum. Using the group composition for the unitary evolution and  letting $\Delta t \to 0$, we can convert the sums into time integrals, such that
\beq
\begin{split}
\label{eq:W_app}
  \chi^{\rm err}_{\alpha\beta}(t,t_0) &\approx \chi_{\alpha\beta}^I + \sum_{p q} G_{p q}\!
	\int_{t_0}^{t}\!\!\!{\rm d} t'\!\left[\mathbb{W}^\dag\!( t') \mathbb{B}^{pq} \mathbb{W}( t')\right]_{\alpha\beta} \,, \\
  \mathbb{W}_{\alpha\beta}( t') &= \text{Tr}\left\{E_\alpha^\dag U(t_0+t',t_0) E_\beta U^\dag\!(t_0+t',t_0) \right\},
  \end{split}
\eeq
making it a linear function of the Lindblad matrix $G$. As a result, the density matrix for the evolution of an initial state $\rho_s$ also becomes a simple linear function of $G$. 

Once we have obtained the full time evolution in terms of the weak-noise error process matrix, we can apply it to the ML-LQT~\eqref{eq:lin_ML} or CS-LQT~\eqref{CS_LT} discussed in the main text. In both cases, we need to calculate 
 the predicted  measurement probabilities for a triple $( \rho_{0,s},t_i, M_\mu)$. Using the above expression, we find 
\beq
  p_{i,s,\mu} = p^{\rm u}_{i,s,\mu} + \sum_{\alpha,\beta}\Phi_{i,s,\mu}^{\alpha\beta} G_{\alpha\beta}\,
  \eeq
  with the definitions introduced in Eqs.~\eqref{eq:p_U}-\eqref{eq:lin_contrib_noise} of the main text.

Since we  have assumed to know the gate unitaries by design, and the measurement configurations are experimentally predefined, the $p^{\rm u}$ probabilities and the $\Phi$ matrix should be calculated  only once. Afterwards, we choose an estimator for $G\mapsto\hat{G}$, which will be obtained by fitting the modelled values of $p_{s,i,\mu}$ to observed frequencies~$f_{s,i,\mu}$. For instance, the  log-likelihood cost function is now explicitly written as
\begin{equation}
  \mathsf{C}_{\rm lin}(G)  =
  -\sum_{i, s, \mu}f_{i,s,\mu}\log{\left( p^{\rm u}_{i,s,\mu} + \sum_{\alpha,\beta}\Phi_{i,s,\mu}^{\alpha\beta} G_{\alpha\beta} \right)}\,,
  \label{eq:ML_for_G}
\end{equation}
which is the solution presented in the main text~\eqref{eq:linearized_loglikelihood} for linearized ML-LQT , and also appears in the CS-LQT constraint~\eqref{CS_LT} after a least-squares approximation.

\section{\bf Convex gradient descent for linear LQT }
\label{app:DIA_PGD}

In this Appendix,  we adapt some of the  efficient  methods for  convex quantum state tomography to our linearized ML-LQT . In particular, we consider the  {diluted iterative algorithm} (DIA)~\cite{Rehacek2007} and the projected gradient descent with momentum (pGDM)~\cite{Bolduc2017} in the context of Lindbladian tomography.

\addtocontents{toc}{\setcounter{tocdepth}{-10}}
\subsection{Diluted iterative algorithm}
\label{eq:DIA_CG_app}

Let us consider the DIA~\cite{Rehacek2007} for our convex problem of linearized ML-LQT , which we rewrite here for convenience
\begin{equation}
  \begin{aligned}
    & \texttt{minimize }\mathsf{C}_{\rm lin}(G) =
    -\sum_{k}f_{k}\log{\left( p^{\rm u}_{k} + \text{Tr}\left\{\Phi^T_{k}G\right\} \right)} \\
    & \texttt{subject to } G = L^{\phantom{\dag}}_{G}L_{G}^\dag\,,
    \label{eq:ML_for_G_2}
  \end{aligned}
\end{equation}
where we recall that the multi-index $k=(s,i,\mu)$ contains all information about the initial state, evolution time, and POVM element, $L_{G}$ is a lower-triangular matrix, and we have simplified the  notation further  by using matrix products.
In contrast to process-QT, $G$ is not constrained further to have unit trace. The DIA algorithm starts by deriving a closed analytical expression for the gradient by varying $\mathsf{C}$ with respect to $G$, which, for our linearized ML-LQT estimator, gives 
\begin{equation}
  \delta\mathsf{C}_{\rm lin}(G)
= -\sum_{a}\frac{f_{k}}{p^{\rm u}_{k} +
  \text{tr}\{\Phi^T_{k}G\}}\text{Tr}\left\{\Phi^T_{k}\delta\!G\right\} =
  \text{Tr}\left\{R\,\delta\!G\right\}\,,
\label{eq:lambda_variation}
\end{equation}
where we have introduced the matrices
\begin{align}
  &R = -\sum_{k}(f_{k}\Phi^T_{k})\big(p^{\rm u}_{k} + \text{Tr}\{\Phi^T_{k}G\}\big)^{-1}\,,\\
  &\delta\!G = \delta L^{\phantom{\dagger}}_{G}L_{G}^\dag + L_{G}^{\phantom{\dagger}}\delta L_{G}^\dag\,.
\end{align}
We note that $R$ is hermitian, since $\Phi_a$ is hermitian.
Rewriting the variation in a ``vector-of-matrices''
form we get
\begin{equation}
  \label{eq:lambda_derivations}
    \delta\mathsf{C}_{\rm lin} 
    = \text{Tr}\left\{
    (L_{G}^\dag R, RL_{G})\cdot(\delta L_{G}, \delta L_{G}^\dag)^T
    \right\} = \langle \boldsymbol{g}_{\mathsf{C}}, \delta \boldsymbol{z}\rangle\,,
\end{equation}
with vectors defined as follows
\begin{equation}
  \boldsymbol{g}_{\rm DIA} = \left(RL, L^\dag R\right)^T,\; \boldsymbol{z}= \left(L^{\phantom{\dag}}_{G}, L_{G}^\dag\right)^T\,.
\end{equation}
We note that the vector $  \boldsymbol{g}_{\mathsf{C}}$ is the {gradient} of the estimator $\mathsf{C}_{\rm lin}$ with respect to $L_G^{\phantom{\dagger}},L^{\dagger}_G$ and, thus,
the  minimum is found when
\begin{equation}
  \boldsymbol{g}_{\rm DIA}= \left(\pd{\mathsf{C}_{\rm lin}}{L^{\phantom{\dag}}_{G}}, \pd{\mathsf{C}_{\rm lin}}{L_{G}^\dag}\right)^{\!\!T} = \boldsymbol{0}
  \quad\Rightarrow\quad RL^{\phantom{\dag}}_{G} = L_{G}^\dag R = 0\,.
\end{equation}
By setting $\delta \boldsymbol{z} = -\eta \boldsymbol{g}_{\rm DIA}$, the estimator
$\mathsf{C}_{\rm lin}$ follows the {gradient descent} to the global minimum.
One can perform different standard line-search techniques
to choose an optimal value of $\eta$, which decreases $\mathsf{C}_{\rm lin}$
as fast as possible at each step. For a chosen $\eta_n$,
the update of the $G$ matrix is given by
\begin{equation}
  G_{n+1} = (C_n + \delta L^{\phantom{\dag}}_{G,n})(C^\dag_n + \delta L^\dag_{G,n})
    = \left(1 - \eta_n R_n\right)G_n\left(1 - \eta_n R_n\right).
\end{equation}

The gradient descent is known to be a sub-optimal optimization algorithm
since it possesses no information about the previous directions of the descent,
and thus can exhibit zig-zag trajectories for problems with large condition
numbers.
One of the most efficient deterministic algorithms circumventing this deficiency
is Conjugate Gradient (CG) \cite{Shewchuk1994}.
Using the recommendation from \cite{Teo2013}, we employ the Polak–Ribière
type of nonlinear CG for the solution of Eq.~\eqref{eq:ML_for_G_2}.

\vspace{2ex}

\begin{algorithmic}[1]

  \REQUIRE $G_{init} = G_0 \rightarrow \boldsymbol{z}_0 = \big(L^{\phantom{\dag}}_{G,0}, L_{G,0}^\dag\big),\,\xi \in (0,1),\, \epsilon$;
  \vspace{1ex}
  \STATE $n = 0$, compute $R_0$;
    \vspace{0.5ex}
  \STATE $ \boldsymbol{g}_{{\rm DIA},0} = \big(R_0L_{G,0}, L_{G,0}^\dag R_0\big),\, \boldsymbol{h}_0 = - \boldsymbol{g}_{{\rm DIA},0}$;
  \vspace{1ex}
  \WHILE{$ \left|\,\text{Tr}\{R_nL_{G,n}\}\,\right| > \epsilon $}
    \vspace{0.5ex}
    \STATE $n = n + 1$;
      \vspace{0.5ex}
    \STATE line search of $\eta_n$ for $\boldsymbol{z}_n = \boldsymbol{z}_{n-1} + \eta_n \boldsymbol{h}_{n-1}$;
      \vspace{0.5ex}
    \STATE compute $R_n \big(\boldsymbol{z}_n\big)$;
      \vspace{0.5ex}
    \STATE $ \boldsymbol{g}_{{\rm DIA},n} = \big(R_nL_{G,n}, L_{G,n}^\dag R_n\big)$;
      \vspace{0.5ex}
    \STATE set $\gamma_n =
    \max{\left\{\dfrac{\langle  \boldsymbol{g}_{{\rm DIA},n}, \boldsymbol{g}_{\mathsf{C},n} -
    \xi \boldsymbol{g}_{{\rm DIA},n-1}\rangle}{\langle  \boldsymbol{g}_{{\rm DIA},n-1}, \ \boldsymbol{g}_{{\rm DIA},n-1}\rangle},
      0\right\}}$ 
        \vspace{0.5ex}
    \STATE $\boldsymbol{h}_n = - \boldsymbol{g}_{{\rm DIA},n} + \gamma_n\boldsymbol{h}_{n-1}$;
      \vspace{0.5ex}
  \ENDWHILE
   \vspace{1ex}
  \RETURN $\hat{\boldsymbol{z}} = (L^{\phantom{\dag}}_{G,n}, L_{G,n}^\dag) \rightarrow \hat{G} = L^{\phantom{\dag}}_{G,n}L_{G,n}^\dag$\,.
   
\end{algorithmic}
\vspace{1ex}

The line search for $\eta_n$ at  step 5 can be implemented by 
choosing arbitrary $\eta'$, $\eta''$, and calculating
$\boldsymbol{z}_n' = \boldsymbol{z}_{n-1} + \eta_n' \boldsymbol{h}_{n-1}$ and
$\boldsymbol{z}_n'' = \boldsymbol{z}_{n-1} + \eta_n'' \boldsymbol{h}_{n-1}$. Then, using the
three values
$\mathsf{C}_{\rm lin}\big(\boldsymbol{z}_{n-1}\big) = \mathsf{C}_{\rm lin}(\eta = 0),\,
\mathsf{C}_{\rm lin}\big(\boldsymbol{z}_n'\big) = \mathsf{C}_{\rm lin}(\eta = \eta_n'),\,
\mathsf{C}_{\rm lin}\big(\boldsymbol{z}_n''\big) = \mathsf{C}_{\rm lin}(\eta = \eta_n'')$,
one interpolates a quadratic polynomial $\mathsf{C}_{\rm lin}(\eta) = a\eta^2 + b\eta + c$,
and uses its minimum as the $\eta_n$. One should be careful though not to
use too large values of $\eta',\,\eta''$, because it could result in breaking of the
first order approximation in $||G||\Delta t\ll 1$ underlying the linearization. We also note that step 8 introduces the so-called the Polak–Ribière factor for the CG descent.
In this work we chose $\eta'' = 2\eta'$ and optimized the value of $\eta'$ for each figure featuring the DIA algorithm from the 
range $(0.1, 1)$. $\xi$ was fixed at $0.5$.

\subsection{Projected  gradient descent with momentum}
\label{eq:pCGM_CG_app}

In this part of the Appendix, we adapt a different type of methods to deal with the positive-semidefinite constraint in the linearized ML-LQT. In state-QT, the
use of a Choleski decomposition to explicitly deal with the constraints of the density matrix  has been shown
to
be responsible for a convergence slowdown of QT algorithms~\cite{Goncalves2016, Shang2017, Bolduc2017}, particularly   when the state $\rho$ to be estimated approaches a pure state and thus has a very small rank.
These works use a method to speed up the estimation by exploiting
a projected descent, such that the density matrix is not constrained to be positive-semidefinite along the descent
(it is just required to remain hermitian), but it is instead projected back to the physical space after
each ML descent iteration.
A suggested algorithm in~\cite{Bolduc2017} is the 
projected gradient descent with momentum (pGDM),
which showed superior convergence  for  QT of high-purity and thus low-rank  states.

The pGDM for
the linearized ML-LQT can be presented as follows. We denote by
$\bar{G}$ and $G$  physical positive-semidefinite
and possibly unphysical matrices, respectively. Since we do not use the Cholesky decomposition, the gradient of the linear estimator is found by varying with respect to $G$ directly
\begin{equation}
  \delta\mathsf{C}_{\rm lin}(G)
= 
  \text{Tr}\left\{R\,\delta\!G\right\}=\langle g_{pGDM},\delta G\rangle\,,
\label{eq:lambda_variation}
\end{equation}
 such that $g_{pGDM}=R$.  The syntax of the algorithm is much simpler in this case

\vspace{2ex}
\begin{algorithmic}[1]
  \REQUIRE $G_{init},\, 0.9 < \gamma < 1,\, \eta > 0,\, \epsilon$;
  \vspace{0.5ex}
  \STATE $n = 0,\, \bar{G}_0 = P\{G_{init}\}, M_0 = 0$;
    \vspace{0.5ex}
  \WHILE{$\left|\mathsf{C}_{\rm lin}(\bar{G}_n) - \mathsf{C}_{\rm lin}(\bar{G}_{n-1})\right| > \epsilon $}
  \vspace{0.5ex}
    \STATE $n = n + 1$;
    \vspace{0.5ex}
    \STATE $M_n = \gamma M_{n-1} - \eta g_{pGDM}(\bar{G}_{n-1})$; 
    \vspace{0.5ex}
    \STATE $\bar{G}_n = P\{\bar{G}_{n-1} + M_n\}$;
    \vspace{0.5ex}
  \ENDWHILE
  \RETURN $\bar{G}_n$\,.
\end{algorithmic}
\vspace{1ex}

We note that, in step 4, we introduce a 
so-called momentum increase/change, which  is controlled by the ``friction'' constant $\gamma$
and the accumulation constant $\eta$. As in the DIA of the previous subsection, one should be careful
not to set too large values of $\eta$ that would imply a  breakdown of the first-order approximation underlying the linearization. We note hat, in contrast to  DIA, the gradient descent step $\eta$ is not optimized by using a line search, nor do we apply a conjugate-gradient strategy. Instead, both $\eta$ and the descent directions are  held fixed.
In step 5 of the algorithm, the projection $P$ is done by diagonalizing the corresponding matrix and setting to zero the contribution from all negative eigenvalues. We also note that one could pre-calculate the new momentum update at
the position based on the previously accumulated
inertia, namely use $g_{pGDM}(\bar{G}_{n-1} + \gamma M_{n-1})$ instead
of $g_{pGDM}(\bar{G}_{n-1})$.
This is called the Nesterov update \cite{Nesterov2004}, and it has been shown to be
beneficial in many non-linear minimization problems. Additionally, when  $\mathsf{C}_{\rm lin}\big(\bar{G}_n\big)$  increases above a threshold, one can
 reset the descent momentum to zero \cite{Donoghue2012}.
In this work we found the values of $\gamma = 0.99$, $\eta = 3\times 10^{-4}$, and no Nesterov update to achieve the fastest convergence to the minimum.

\bibliography{main}

\end{document}